\documentclass[twoside,english]{elsarticle}
\usepackage[T1]{fontenc}
\usepackage[utf8]{inputenc}
\usepackage{array}
\usepackage{textcomp}
\usepackage{multirow}
\usepackage{amsmath}
\usepackage{amssymb}
\usepackage{graphicx}
\makeatletter
\providecommand{\tabularnewline}{\\}
\journal{Example: Astroparticle Physics}
\usepackage{babel}
\usepackage{babel}
\usepackage{babel}
\usepackage{babel}
\usepackage{babel}
\usepackage{babel}
\usepackage{babel}
\makeatother
\usepackage{babel}
\begin{document}
\title{Ill-posed formulation of the emission source localization in the
radio-detection experiments of extensive air showers}
\author[rvt]{Ahmed~Rebai\corref{cor1}}
\ead{ahmed.rebai@subatech.in2p3.fr}
\author[focal]{Tarek~Salhi}
\ead{tareksalhi@gmail.com}
\author[rvt]{Pascal~Lautridou\corref{cor1}}
\ead{Pascal.Lautridou@subatech.in2p3.fr}
\author[rvt]{Olivier Ravel}
\ead{Olivier.Ravel@subatech.in2p3.fr}
\cortext[cor1]{Corresponding authors}
\address[rvt]{SUBATECH IN2P3-CNRS/Université de Nantes/Ecole des Mines de Nantes, Nantes, France}
\address[focal]{Ecole des Mines de Nantes, Nantes, France }
\begin{abstract}
Reconstruction of the curvatures of radio wavefronts of air showers
initiated by ultra high energy cosmic rays is discussed based on minimization
algorithms commonly used. We emphasize the importance of the convergence
process induced by the minimization of a non-linear least squares
function that affects the results in terms of degeneration of the
solutions and bias. We derive a simple method to obtain a satisfactory
estimate of the location of the main point of emission source, which
mitigates the problems previously encountered.\end{abstract}
\begin{keyword}
UHECR \sep radio-detection \sep antennas \sep non-convex analysis
\sep optimization\sep ill-posed problem. 
\end{keyword}
\maketitle
\section{Introduction}
The determination of the nature of the ultra-high energy cosmic rays
(UHECR) is an old fundamental problem in cosmic rays studies. Numerous
are the difficulties. New promising approaches could emerge from the
use of the radio-detection method which exploits, through antennas,
the radio signal that accompanies the development of the extensive
air shower (EAS). Several experimental prototypes like CODALEMA \cite{key-1}
in France and LOPES \cite{key-2} in Germany shown the feasibility
and the potential of the method to reconstruct EAS parameters, as
the arrival direction, the impact location at ground, the lateral
distribution function of the electric field, or the primary particle
energy \cite{key-3,key-4,key-5,key-6,key-7}. However, the temporal
radio wavefront characteristics remain still poorly determined \cite{key-8,key-8-1},
although its knowledge could be consider as one of the first steps
in retrieving information about the EAS itself. The importance of
this information resides in its potential sensitivity to the nature
of the primary particle, especially because the existence of a curvate
radio wavefront (a spherical wavefront) could provide the location
of the main point of the emission source, and possibly an estimation
of Xmax, event by event. Indeed, the arrival timing being defined
by the maximum amplitude of the radio signal, it is more likely linked
to a limited portion of the longitudinal development of the shower
(and so especially at the point of maximum) \cite{key-8-2}.\medskip{}

\begin{figure}[hbtp]
\centering 
\includegraphics[width=11cm,height=6cm]{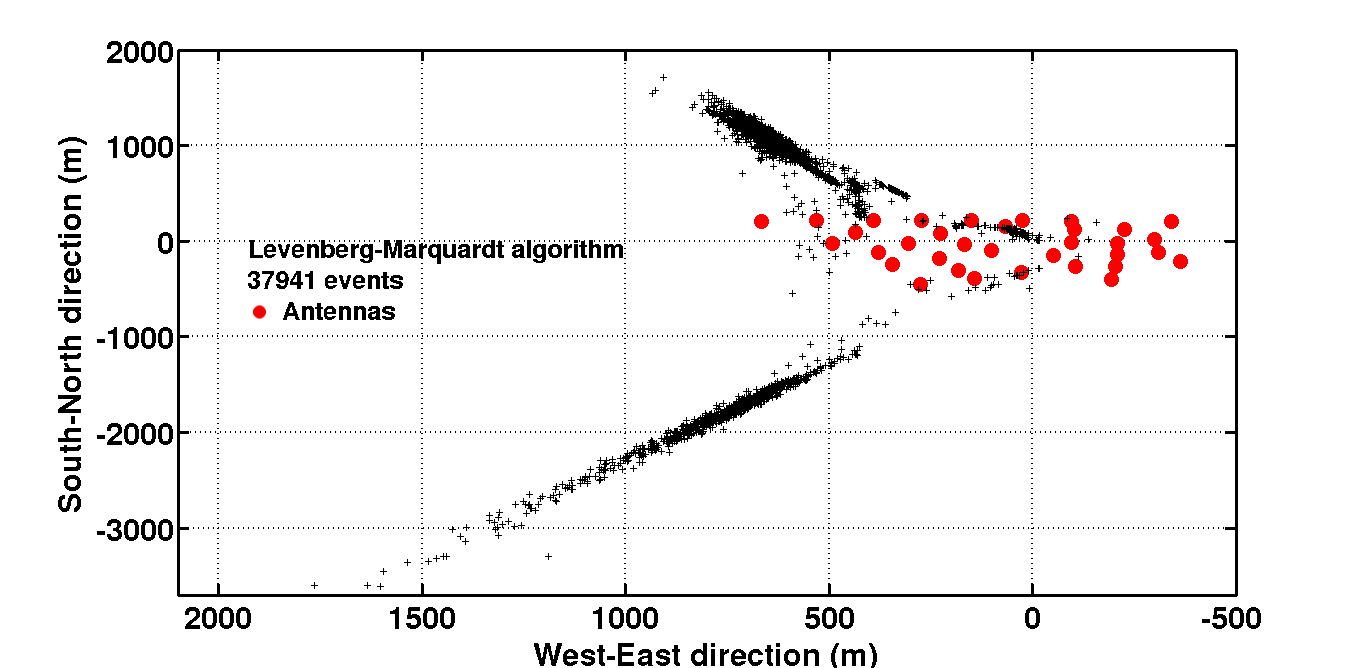}
\caption{Typical result of reconstruction of two entropic emitters at ground,
observed with the stand-alone stations of CODALEMA, through standard
minimization algorithms. Despite the spreading of the reconstructed
positions, these two transmitters are, in reality, two stationary
point sources. }
\label{codanoise} 
\end{figure}

Moreover, the migration of present small scale radio-prototypes to
large scale experiments spread over surfaces of several tens of $1000\: km^{2}$
using self-triggered antennas, is challenging. This technique is subjected
to delicate limitations in regard to UHECR recognition, due to noises
induced by human activities (high voltage power lines, electric transformers,
cars, trains and planes) or by stormy weather conditions (lightning).
Figure \ref{codanoise} shows a typical reconstruction of sources
obtained with the CODALEMA experiment \cite{key-9}, by invoking a
spherical wave minimization. Such patterns are also commonly observed
in others radio experiments \cite{key-9-1,key-9-2}. In most of the
cases, one of the striking results is that these emission sources
are reconstructed with great inaccuracy, although they are fixed and
although the number of measured events is high. By extension, a cosmic
event being a single realization of the detected observables (arrival
time and peak amplitude on each antenna), interpretation of such methods
of reconstruction for the identification of a point source can become
even more delicate, even using statistical approaches.

The commonly used technique relies on the minimization of an objective
function which depends on the assumed shape of the wavefront, using
the arrival times and locations of the antennas. The aim of this paper
is to highlight that the minimization of such an objective function,
incorporating a spherical wave front, can be an ill-posed problem.
We will show that it originates from strong dependencies of the convergence
of the minimization algorithms with initial parameters, from the existence
of degenerations of the solutions (half lines) which can trap most
of the common algorithms, and from the existence of offsets (bias)
in the reconstructed positions. Finally, by avoiding more complex
estimates based on advanced statistical theories, we got to deduce
a simple method to obtain a significant estimate of the source location.
We compared the exact results with our numerical reconstructions performed
on a test array.

\section{Reconstruction with common algorithms}

\begin{figure}[htbp]
\centering
\includegraphics[width=11cm,height=6cm]{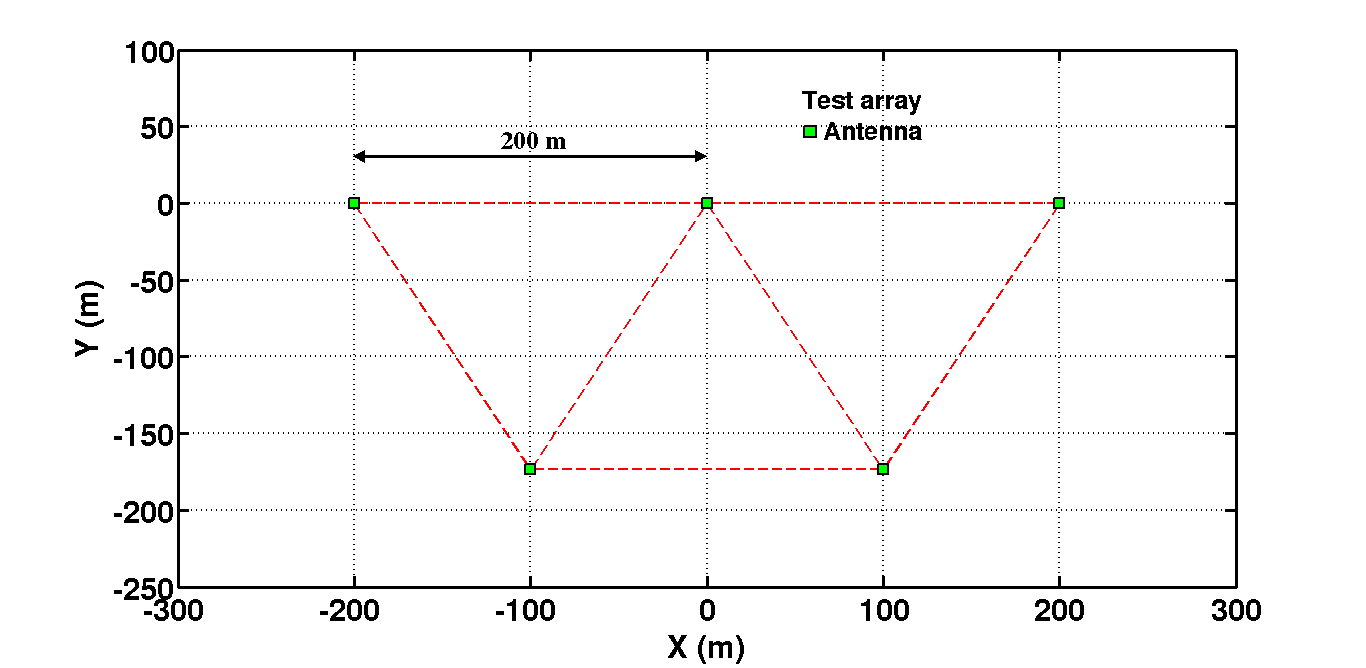}
\caption{Scheme of the antenna array used for the simulations. The antenna
location is took from a uniform distribution of $1\: m$ width. }
\label{how_test_estim} 
\end{figure}

The performances of different algorithms has been tested using the
simplest test array of antennas. Within the constraints imposed by
the number of free parameters used for reconstruction, we choose an
array of 5 antennas for which the antennas positions $\overrightarrow{r_{i}}=\left(x_{i},y_{i},z_{i}\right)$
are fixed (see Fig. \ref{how_test_estim}) (this corresponds to a
multiplicity of antennas similar to that sought at the detection threshold
in current setups).

A source S with a spatial position $\overrightarrow{r_{s}}=\left(x_{s},y_{s},z_{s}\right)$
is set at the desired value. Assuming $t_{s}$ the unknown instant
of the wave emission from S, c the wave velocity in the medium considered
constant during the propagation, and assuming that the emitted wave
is spherical, the reception time $t_{i}$ on each antenna $i\in\left\{ 1,\ldots,N\right\} $
can written: 
\[
t_{i}=t_{s}+\frac{\sqrt{\left(x_{i}-x_{s}\right)^{2}+\left(y_{i}-y_{s}\right)^{2}+\left(z_{i}-z_{s}\right)^{2}}}{c}+G(0,\sigma_{t})
\]
where $G(0,\sigma_{t})$ is the Gaussian probability density function
centered to $t=0$ and of standard deviation $\sigma_{t}$. This latter
parameter stand for the the global time resolution, which depends
as well on technological specifications of the apparatus than on analysis
methods.

The theoretical predictions are compared to the reconstructions given
by the different algorithms. The latter are setup in two steps. First,
a planar adjustment is made, in order to pres-tress the region of
the zenith angle $\theta$ and azimuth angle $\phi$ of the source
arrival direction. It specifies a target region in this subset of
the phase space, reducing the computing time of the search of the
minimum of the objective function of the spherical emission. Reconstruction
of the source location is achieved, choosing an objective-function
that measures the agreement between the data and the model of the
form, by calculating the difference between data and a theoretical
model (in frequentist statistics, the objective-function is conventionally
arranged so that small values represent close agreement):

\begin{equation}
f(\vec{r_{s}},t_{s}^{*})=\frac{1}{2}\sum_{i=1}^{N}\left[\left\Vert \overrightarrow{r_{s}}-\overrightarrow{r_{i}}\right\Vert ^{2}-\left(t_{s}^{*}-t_{i}^{*}\right)^{2}\right]^{2}
\end{equation}

The partial terms $\left\Vert \overrightarrow{r_{s}}-\overrightarrow{r_{i}}\right\Vert ^{2}-\left(t_{s}^{*}-t_{i}^{*}\right)^{2}$
represents the difference between the square of the radius calculated
using coordinates and the square of the radius calculated using wave
propagation time for each of the N antennas. The functional $f$ can
be interpreted as the sum of squared errors. Intuitively the source
positions $\overrightarrow{r_{s}}$ at the instant $t_{s}$ is one
that minimizes this error. 

In the context of this paper, we did not use genetic algorithms or
multivariate analysis methods but we focused on three minimization
algorithms, used extensively in statistical data analysis software
of high energy physics\cite{key-10,key-11}: Simplex, Line-Search
and Levenberg-Marquardt (see table \ref{diff_algo 1}). They can be
found in many scientific libraries as the Optimization Toolbox in
Matlab, the MPFIT in IDL and the library Minuit in Root that uses
2 algorithms Migrad and Simplex which are based respectively on a
variable-metric linear search method with calculation of the objective
function first derivative and a simple search method. For the present
study, we have used with their default parameters.

\begin{table}
\label{diff_algo 1}

\begin{tabular}{|>{\centering}p{22mm}|>{\centering}p{30mm}|>{\centering}p{30mm}|>{\centering}p{30mm}|}
\hline 
Minimization algorithms  & Levenberg-Marquardt  & Simplex  & Line-Search\tabularnewline
\hline 
Libraries  & lsqnonlin - MPFIT  & fminsearch - SIMPLEX  & MIGRAD - lsqcurvefit\tabularnewline
\hline 
Software  & Optimization Toolbox Matlab - IDL  & Optimization Toolbox Matlab - MINUIT-ROOT  & Optimization Toolbox Matlab - MINUIT-ROOT\tabularnewline
\hline 
Method Principles  & Gauss-Newton method combined with trust region method  & Direct search method  & Compute the step-size by optimizing the merit function $f(x+t.d)$ \tabularnewline
\hline 
Used information  & Compute gradient $(\nabla f)_{k}$ and an approximate hessian $(\nabla^{2}f)_{k}$  & No use of numerical or analytical gradients  & $f(x+t.d,\, d)$ where $d$ is a direction descent computed with gradient/hessian\tabularnewline
\hline 
Advantages / Disadvantages  & Stabilize ill-conditioned Hessian matrix / time consuming and local
minimum trap  & No reliable information about parameter errors and correlations  & Need initialization with another method, give the optimal step size
for the optimization algorithm then reduce the complexity\tabularnewline
\hline 
\end{tabular}

\caption{Summary of the different algorithms and methods used to minimize the
objective-function. The second row indicates framework functions corresponding
to each algorithm; third recalls the framework names. The key information
used for optimization are recalled down, noting that a differentiable
optimization algorithm (ie. non-probabilistic and non-heuristic) consists
of building a sequence of points in the phase space as follows: $x_{k+1}=x_{k}+t_{k}.d_{k}$,
and that it is ranked based on its calculation method of $t_{k}$
and $d_{k}$ parameters (see \cite{key-12}).}
\end{table}

\begin{figure}[!h]
\centering
\includegraphics[width=6cm,height=6cm]{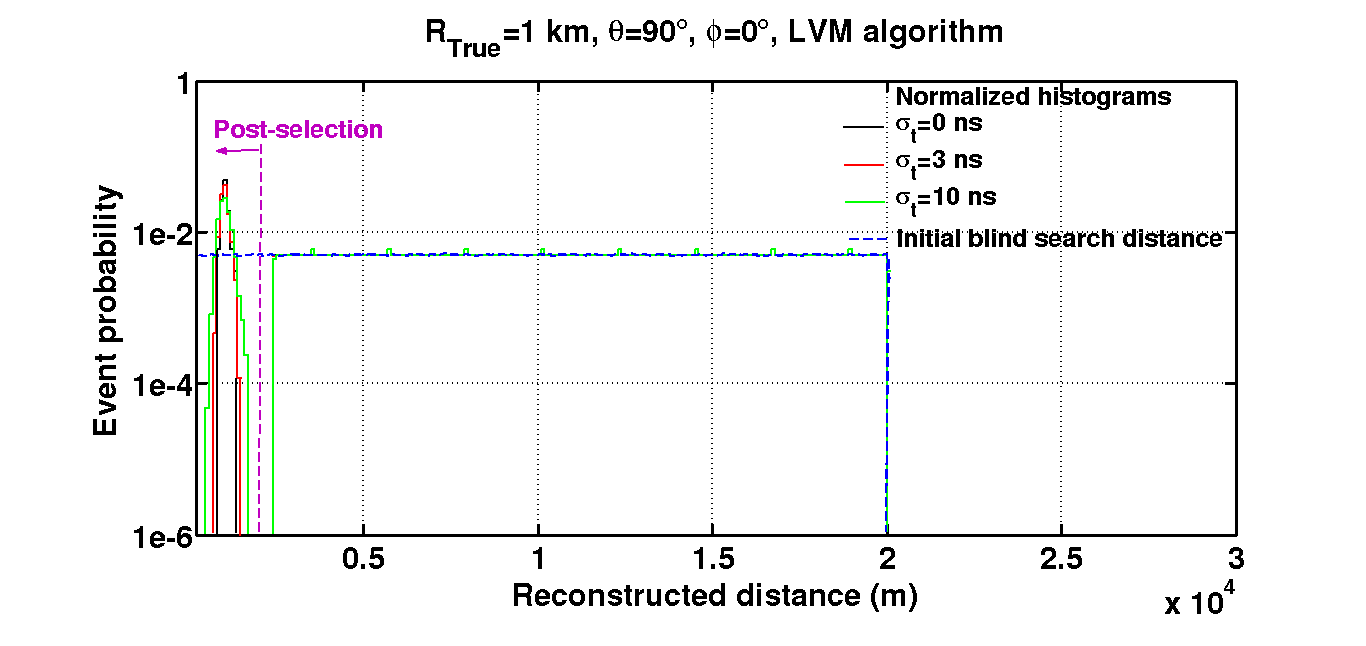}
\includegraphics[width=6cm,height=6cm]{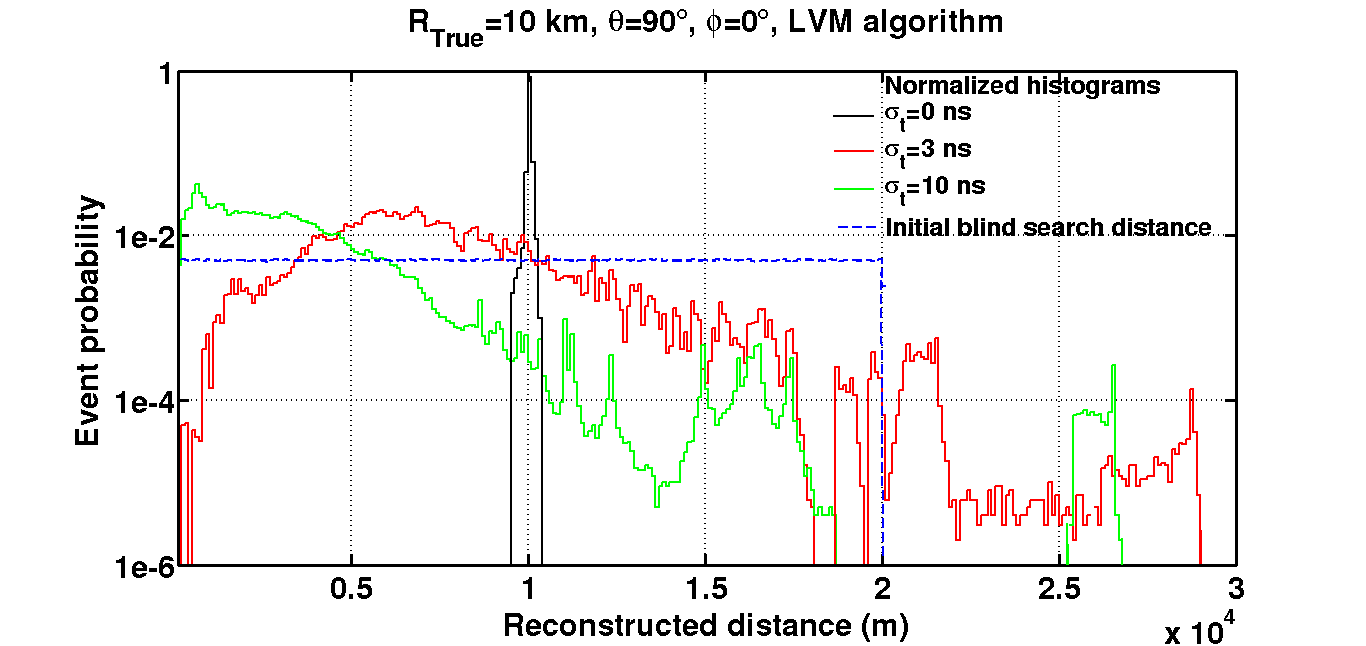}
\caption{Results of the reconstruction of a source with a radius of curvature
equal to 1 and 10 km with the LVM algorithm. For $R_{true}=1\, km$,
the effect of the blind search leads to non-convergence of the LVM
algorithm, when initialization values are greater than $R_{true}=1\, km$.}
\label{LVM_REC_10km} 
\end{figure}

\begin{figure}[!h]
\centering
\includegraphics[width=6cm,height=6cm]{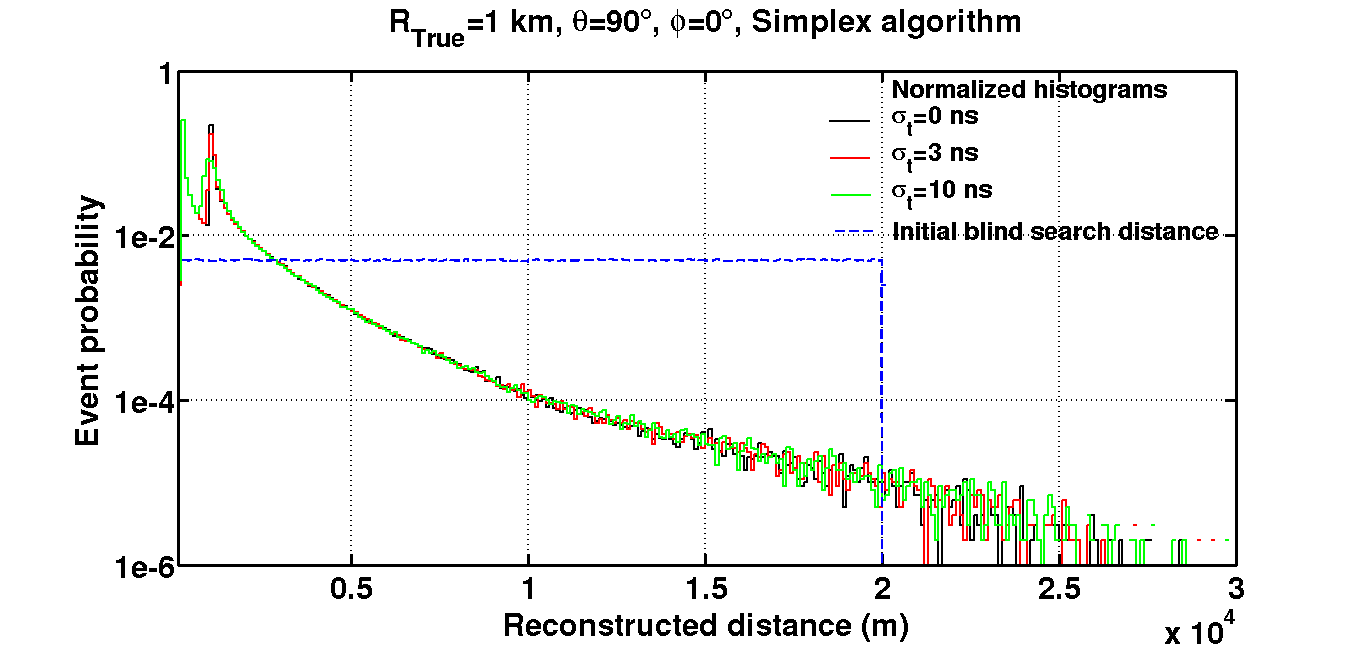}
\includegraphics[width=6cm,height=6cm]{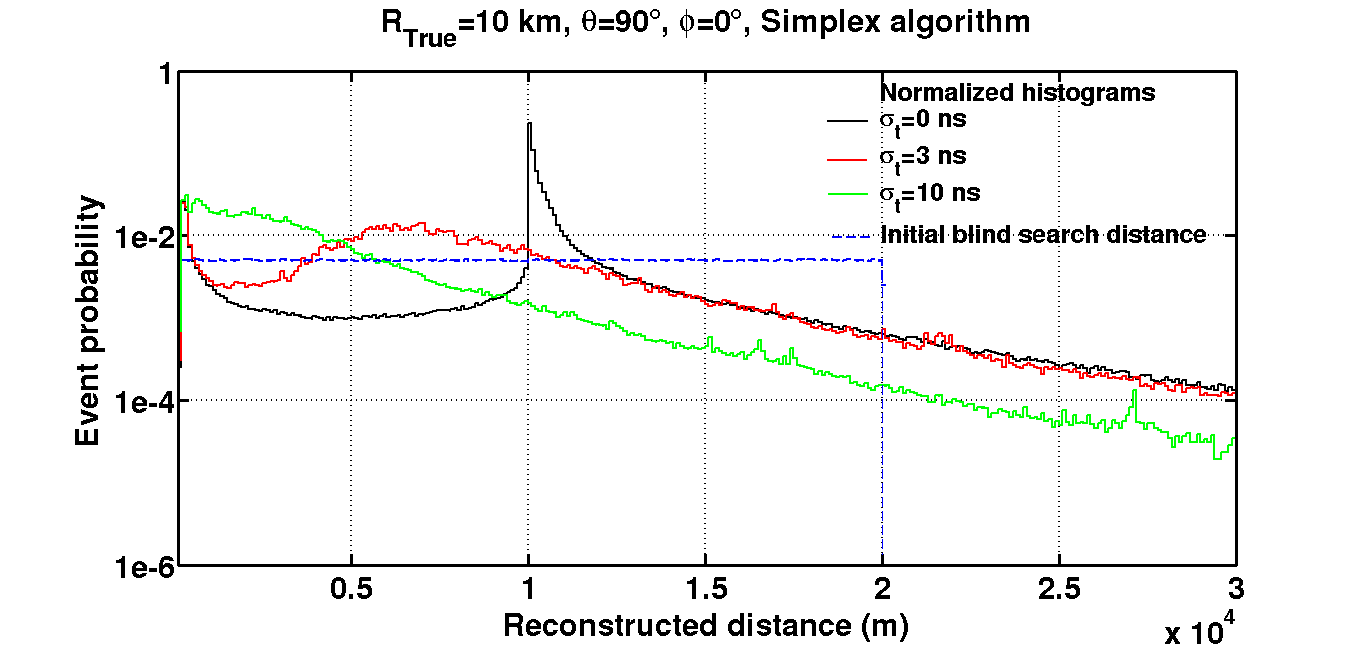}
\caption{Results of the reconstruction of a source with a radius of curvature
equal to 1 and 10 km with the Simplex algorithm.}
\label{Simpl_1_10km} 
\end{figure}

We tested three time resolutions with times values took within $3\,\sigma_{t}$
.
\begin{itemize}
\item $\sigma_{t}=0\, ns$ plays the role of the perfect theoretical detection
and serves as reference; 
\item $\sigma_{t}=3\, ns$ reflects the optimum performances expected in
the current state of the art; 
\item $\sigma_{t}=10\, ns$ stands for the timing resolution estimate of
an experiment like CODALEMA \cite{key-1-1}. 
\end{itemize}
For every source distance and temporal resolution, one million events
were generated. Antenna location was taken in a uniform distribution
of $1\, m$ width. A blind search was simulated using uniform distribution
of the initial $r_{s}$ values from $0.1\, km$ to $20\, km$. Typical
results obtained with our simulations are presented in Figures \ref{LVM_REC_10km}
and \ref{Simpl_1_10km}. The summary of the reconstructed parameters
is given in table \ref{ResulSimul}. 

\begin{figure}
\centering
\includegraphics[width=11cm,height=6cm]{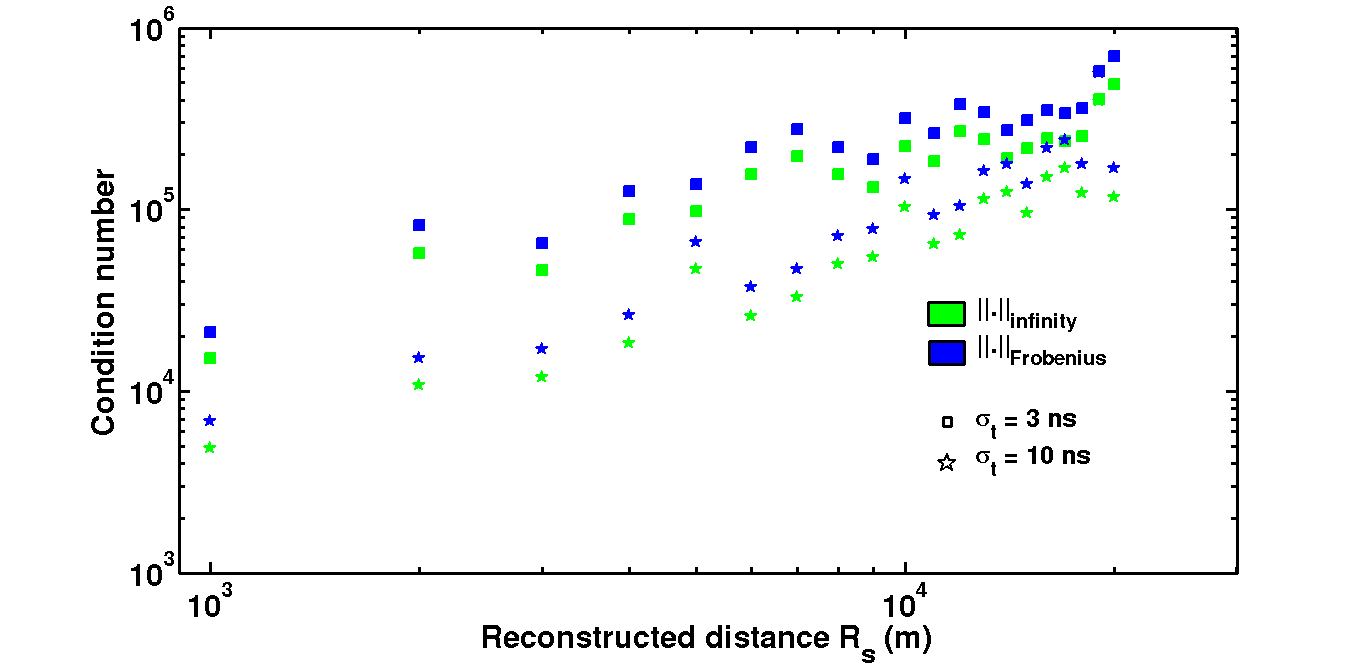}
\label{conditionement}
\caption{Condition numbers obtained using the formula $Cond(Q)=\Vert Q\Vert.\Vert Q^{-1}\Vert$with
$Q$ the Hessian matrix (see next section) as a function of the source
distance and for different timing resolutions. The large values of
conditioning suggest that we face an ill-posed problem. }
\end{figure}

\begin{table}
\label{ResulSimul}%
\begin{tabular}{|c|c|c|c|c|c|}
\hline 
\multicolumn{6}{|c|}{Reconstruction Results}\tabularnewline
\hline 
$\sigma_{t}\,(ns)$  & $R_{true}(m)$  & $Algorithms$ & \multicolumn{1}{c|}{$R_{mean}(m)$} & \multicolumn{1}{c|}{$R_{mode}(m)$} & \multicolumn{1}{c|}{$\sigma^{R}(m)$}\tabularnewline
\hline 
\multirow{6}{*}{0} & \multirow{2}{*}{1000 } & Levenberg-Marquardt  & (10071) 1002  & (1081) 1081 & (5763) 102\tabularnewline
\cline{3-6} 
 &  & Simplex  & \multicolumn{1}{c|}{1198} & \multicolumn{1}{c|}{1133} & \multicolumn{1}{c|}{1477}\tabularnewline
\cline{2-6} 
 & \multirow{2}{*}{3000} & Levenberg-Marquardt  & (9960) 3082  & (2998) 2998  & (5781) 302\tabularnewline
\cline{3-6} 
 &  & Simplex  & \multicolumn{1}{c|}{3134} & \multicolumn{1}{c|}{3272} & \multicolumn{1}{c|}{3437}\tabularnewline
\cline{2-6} 
 & \multirow{2}{*}{10000} & Levenberg-Marquardt  & \multicolumn{1}{c|}{9999} & \multicolumn{1}{c|}{9997} & \multicolumn{1}{c|}{56}\tabularnewline
\cline{3-6} 
 &  & Simplex  & \multicolumn{1}{c|}{10466} & \multicolumn{1}{c|}{9929} & \multicolumn{1}{c|}{5817}\tabularnewline
\hline 
\multirow{6}{*}{3 } & \multirow{2}{*}{1000} & Levenberg-Marquardt  & (10071) 1003  & (934) 934  & (5763) 108\tabularnewline
\cline{3-6} 
 &  & Simplex  & \multicolumn{1}{c|}{1199} & \multicolumn{1}{c|}{168} & \multicolumn{1}{c|}{1486}\tabularnewline
\cline{2-6} 
 & \multirow{2}{*}{3000} & Levenberg-Marquardt  & (9954) 3068  & (2874) 2874  & (5792) 495\tabularnewline
\cline{3-6} 
 &  & Simplex  & \multicolumn{1}{c|}{3132} & \multicolumn{1}{c|}{3010} & \multicolumn{1}{c|}{3485}\tabularnewline
\cline{2-6} 
 & \multirow{2}{*}{10000} & Levenberg-Marquardt  & \multicolumn{1}{c|}{7174} & \multicolumn{1}{c|}{6877} & \multicolumn{1}{c|}{3021}\tabularnewline
\cline{3-6} 
 &  & Simplex  & \multicolumn{1}{c|}{8194} & \multicolumn{1}{c|}{6479} & \multicolumn{1}{c|}{6154}\tabularnewline
\hline 
\multirow{6}{*}{10} & \multirow{2}{*}{1000} & Levenberg-Marquardt  & (10068) 985  & (964) 964  & (5767) 175\tabularnewline
\cline{3-6} 
 &  & Simplex  & \multicolumn{1}{c|}{1189} & \multicolumn{1}{c|}{199} & \multicolumn{1}{c|}{1507}\tabularnewline
\cline{2-6} 
 & \multirow{2}{*}{3000} & Levenberg-Marquardt  & (9703) 2238  & (1620) 1620  & (6125) 877\tabularnewline
\cline{3-6} 
 &  & Simplex  & \multicolumn{1}{c|}{2760} & \multicolumn{1}{c|}{2070} & \multicolumn{1}{c|}{3703}\tabularnewline
\cline{2-6} 
 & \multirow{2}{*}{10000} & Levenberg-Marquardt  & \multicolumn{1}{c|}{2770} & \multicolumn{1}{c|}{667} & \multicolumn{1}{c|}{2305}\tabularnewline
\cline{3-6} 
 &  & Simplex  & \multicolumn{1}{c|}{3675} & \multicolumn{1}{c|}{934} & \multicolumn{1}{c|}{4048}\tabularnewline
\hline 
\end{tabular}
\centering
\caption{Summary of parameters reconstructed with different algorithms for
several distances of source and several timing resolutions. On the
Levenberg-Marquardt, the results in parentheses are those taking into
account the flat portion of the resulting distribution (see Fig. \ref{LVM_REC_10km}).
They are typical of initialization values which are starting too far
from the actual source distance. The Line-Search method was ultimately
rejected for this quantitative study, because results too dependent
on the starting algorithm fixing the initial conditions.}
\end{table}
Whatever the simulations samples (versus any source distances, arrival
directions, time resolutions), (also with several detector configurations)
and the three minimization algorithms, large spreads were generally
observed for the source locations reconstructed. This suggests that
the objective-function presents local minima. Moreover, the results
depend strongly on initial conditions. All these phenomena may indicate
that we are facing an ill-posed problem. Indeed, condition number
calculations \cite{key-14} (see Fig. \ref{conditionement}), which
measures the sensitivity of the solution to errors in the data (as
the distance of the source, the timing resolution, etc.), indicate
large values ($>10^{4}$), when a well-posed problem should induce
values close to 1. 

To understand the observed source reconstruction patterns, we have
undertaken to study the main features of this objective-function.

\section{Study of the objective-function for the spherical emission}
To estimate the source position $X_{s}=(\vec{r_{s}},t_{s})$ using
the sequence of arrival $t_{i}$, the natural method is to formulate
an unconstrained optimization problem of type a non-linear least square \cite{key-12}, starting from eq. 1. which can rewrite%
\footnote{In practice of the minimization, it is usual to take into account
errors on the measured parameters by putting them in the objective
function denominator. In our theoretical study, it is assumed that
the arrival times errors are the same for all the antennas ($\sigma_{t}=constant~\forall~i$).
The present studied functional is generic and does not include errors,
but as will see later, introduction of a multiplicative constant doesn't
change the results of our study.%
} (see the notations listed in table \ref{table_nomenclature}): 

\begin{equation}
f\left(X_{s}\right)=\frac{1}{2}\sum_{i=1}^{N}\left[\left\Vert \overrightarrow{r_{s}}-\overrightarrow{r_{i}}\right\Vert _{2}^{2}-\left(t_{s}^{*}-t_{i}^{*}\right)^{2}\right]^{2}=\frac{1}{2}\sum_{i=1}^{N}f_{i}^{2}(X_{s})
\end{equation}

\begin{table}
\begin{centering}
\begin{tabular}{|>{\centering}p{15cm}|}
\hline 
$\vec{r_{s}}$, $\vec{r_{i}}$: position of the source, position of
the $i^{th}$ antenna \tabularnewline
$t_{s}$, $t_{i}$: emission time of the signal, signal arrival time
at the $i^{th}$ antenna \tabularnewline
$t_{s}^{*}$, $t_{i}^{*}$: reduced time variables (ie. $t^{*}=c.t$)\tabularnewline
$\sigma_{i}^{t}$: time resolution on the $i^{th}$ antenna \tabularnewline
$X_{s}$, $X_{i}$ spacio-temporal position of the source, of the
$i^{th}$ antenna \tabularnewline
$\nabla f$, $\nabla^{2}f$: first and second derivative of the objective
function $f$\tabularnewline
$M=I_{4}-2E_{44}=\begin{bmatrix}1 & 0 & 0 & 0\\
0 & 1 & 0 & 0\\
0 & 0 & 1 & 0\\
0 & 0 & 0 & -1
\end{bmatrix}$ : second order tensor related to the Minkowski metric\tabularnewline
$Q$, $L_{i}$: quadratic and linear form \tabularnewline
$<.|.>$: inner product \tabularnewline
$X^{T}$: transpose of a vector or a matrix X\tabularnewline
\hline 
\end{tabular}\caption{List of notations}

\par\end{centering}

\centering{}\label{table_nomenclature} 
\end{table}

Several properties of the objective-function $f$ were studied: the
coercive property to indicate the existence of at least one minima,
the non-convexity to indicate the existence of several local minima,
and the jacobian to locate the critical points. (Bias study, which
corresponds to a systematic shift of the estimator, is postponed to
another contribution). In mathematical terms, this analysis amounts
to: 
\begin{itemize}
\item Estimate the limits of $f$ to make evidence of critical points; obviously,
the objective function $f$ is positive, regular and coercive. Indeed,
$f$ tends to $+\infty$ when $\|X\|\to\pm\infty$, because it is
a polynomial and contains positive square terms. So, $f$ admits at
least a minimum.
\item Verify the second optimality condition: the convexity property of
a function on a domain for a sufficiently regular function is equivalent
to positive-definiteness character of its Hessian matrix. 
\item Solve the first optimality condition: $\nabla f(X_{s})=0$ (jacobian)
to find the critical points. 
\end{itemize}

\subsection{Convexity property}

Using $f_{i}(X_{s})=(X_{s}-X_{i})^{T}.M.(X_{s}-X_{i})$ where $M$
designates the \textit{Minkowski} matrix and given $\nabla f_{i}(X_{s})=2.M(X_{s}-X_{i})$,
the $f$ gradient function can written (see appendix 1): 
\[
\frac{1}{2}\nabla f(X_{s})=(\sum f_{i}(X_{s}))M.X_{s}-M.(\sum f_{i}(X_{s})X_{i})
\]
The Hessian matrix, which is the $f$ second derivative can written:
\[
\nabla^{2}f(X_{s})=\sum\nabla f_{i}(X_{s}).\nabla f_{i}^{T}+\sum f_{i}.\nabla^{2}f_{i}
\]
that becomes, replacing $\nabla f_{i}$ by its expression: 
\[
\nabla^{2}f(X_{s})=(\sum f_{i}(X_{s})).M+2M.[N.X_{s}.X_{s}^{T}+\sum X_{i}X_{i}^{T}-X_{s}(\sum X_{i})^{T}-(\sum X_{i})X_{s}^{T}].M
\]
Using a Taylor series expansion to order $2$ (see appendix $1$),
an expanded form of the Hessian matrix, equivalent to the previous
formula of the $f$ second derivative, is:

{\tiny 
\begin{equation}
\frac{1}{2}Q\left(X_{s},X_{i}\right)=\left[\begin{array}{cccc}
\sum_{i}K_{i}+2\sum_{i}\left(x_{s}-x_{i}\right)^{2} & 2\sum_{i}\left(x_{s}-x_{i}\right)\left(y_{s}-y_{i}\right) & 2\sum_{i}\left(x_{s}-x_{i}\right)\left(z_{s}-z_{i}\right) & 2\sum_{i}\left(x_{s}-x_{i}\right)\left(t_{i}^{*}-t_{s}^{*}\right)\\
* & \sum_{i}K_{i}+2\sum_{i}\left(y_{s}-y_{i}\right)^{2} & 2\sum_{i}\left(y_{s}-y_{i}\right)\left(z_{s}-z_{i}\right) & 2\sum_{i}\left(y_{s}-y_{i}\right)\left(t_{i}^{*}-t_{s}^{*}\right)\\
* & * & \sum_{i}K_{i}+2\sum_{i}\left(z_{s}-z_{i}\right)^{2} & 2\sum_{i}\left(z_{s}-z_{i}\right)\left(t_{i}^{*}-t_{s}^{*}\right)\\
* & * & * & -\sum_{i}K_{i}+2\sum_{i}\left(t_{i}^{*}-t_{s}^{*}\right)^{2}
\end{array}\right]
\end{equation}
}{\tiny \par}

This latter allowed us to study the convexity of $f$ (see appendix
1). Indeed, because its mathematical form is not appropriate for a
direct use of the convexity definition, we have preferred to use the
property of semi-positive-definiteness of the Hessian matrix. Our
calculus lead to the conclusion that: 
\begin{itemize}
\item Using the criterion of Sylvester \cite{key-13} and the analysis
of the principal minors of the Hessian matrix , we find that $f$
is not convex on small domains, and thus is likely to exhibit several
local minima, according to $X_{s}$ and $X_{i}$. It is these minima,
which induce convergence problems to the correct solution for the
common minimization algorithms. 
\end{itemize}

\subsection{Critical points}

The study of the first optimality condition (Jacobian = 0) gives the
following system $\nabla f(X_{s})=0$ and allows finding the critical
points and their phase-space distributions. Taking into account the
following expression:

\begin{center}
$\frac{1}{2}\nabla f(\bar{X_{s}})=(\sum f_{i}(\bar{X_{s}}))M.\bar{X_{s}}-M.(\sum f_{i}(\bar{X_{s}})X_{i})$ 
\par\end{center}

we get the relation: 

\begin{equation}
\overline{X}_{s}=\sum_{i=1}^{N}\frac{f_{i}(\overline{X}_{s})}{\sum_{j}f_{j}(\overline{X}_{s})}\, X_{i}
\end{equation}

This formula looks like the traditional relationship of a barycenter.
Thus, we interpret it in terms of the antennas positions barycenter
and its weights. The weight function $f_{i}$ expressing the space-time
distance error between the position exact and calculated, the predominant
direction will be the one presenting the greatest error between its
exact and calculated position. The antennas of greatest weight will
be those the closest to the source. 

In practice (see appendix 2), because the analytical development of
this optimality condition in a three-dimensional formulation is not
practical, especially considering the nonlinear terms, we chose to
study particular cases. We considered the case of a linear antennas
array (1D) for which the optimality condition is easier to express
with an emission source located in the same plane. This approach allows
us to understand the origin of the observed degeneration which appears
from the wave equation invariance by translation and by time reversal
(known reversibility of the wave equation in theory of partial differential
equations) and provides us a intuition of the overall solution. It
also enlightens the importance of the position of the actual source
relative to the antennas array (the latter point is linked to the
convex hull of the antenna array and is the object of the next section).
Our study led to the following interpretations:
\begin{itemize}
\item The iso-barycenter of the antenna array (of the lit antennas for a
given event) plays an important role in explaining the observed numerical
degeneration. The nature of the critical points set determines the
convergence of algorithms and therefore the reconstruction result. 
\item There are strong indications, in agreement with the experimental results
and our calculations (for 1D geometry), that the critical points are
distributed on a line connecting the barycenter of the lit antennas
and the actual source location. We used this observation to construct
an alternative method of locating the source (section 4). 
\item According to the source position relative to the antenna array, the
reconstruction can lead to an ill-posed or well-posed problem, in
the sense of J. Hadamard.
\end{itemize}

\subsection{Convex hull}

In the previous section we pointed that to face a well-posed problem
(no degeneration in solution set), it was necessary to add constraints
reflecting the propagation law in the medium, the causality constraints,
and a condition linking the source location and the antenna array,
the latter inducing the concept of convex hull of the array of antennas. 
From appendix 2, we also saw that analytically the critical points
evidence could become very complex from the mathematical point of
view. Therefore, we chose again an intuitive approach to characterize
the convex hull, by exploring mathematically the case of a linear
array with an emission source located in the same plane. This is the
subject of the appendix 3. \medskip{}

\begin{figure}
\centering
\includegraphics[width=11cm,height=7cm]{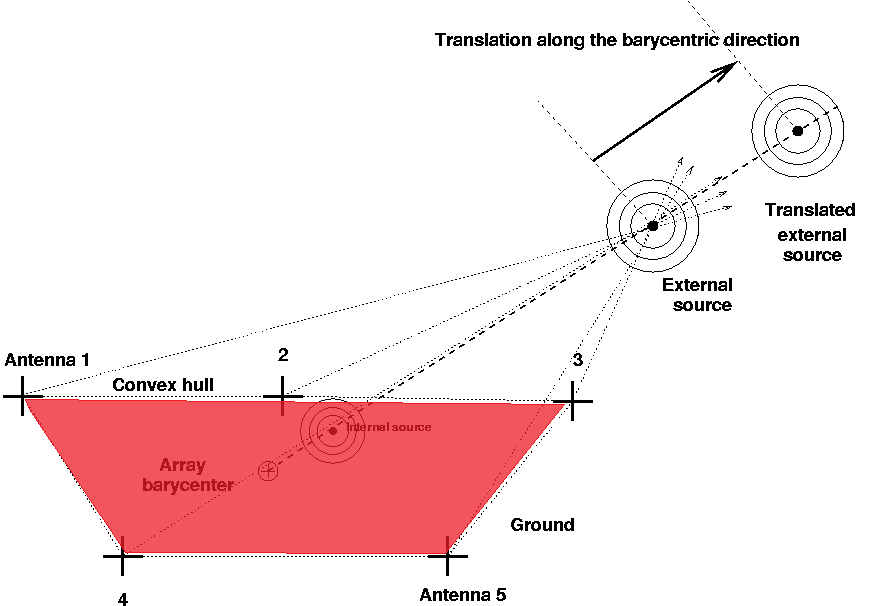}
\caption{Scheme of the reconstruction problem of spherical waves for our testing
array of antennas (2D), with a source located at ground. For this
configuration, the convex hull becomes the surface depicted in red.
The result is the same for a source in the sky. }
\label{Det2D}
\end{figure}

The results extend to a 2D antenna array, illuminated by a source
located anywhere at ground, arguing that it is possible to separate
the array into sub-arrays arranged linearly. The superposition of
all the convex segments of the sub-arrays leads then to conceptualize
a final convex surface, built by all the peripheral antennas illuminated
(see Figure \ref{Det2D}). 

\begin{figure}
\centering
\includegraphics[width=6cm,height=6cm]{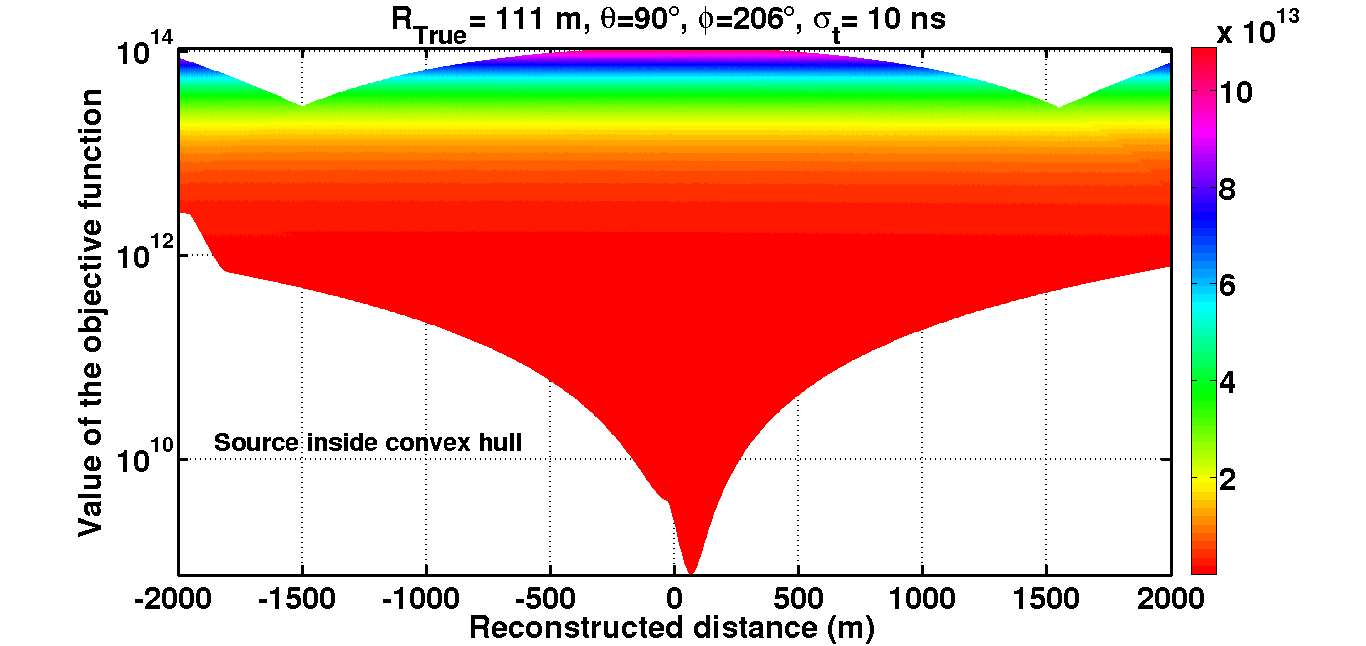}
\includegraphics[width=6cm,height=6cm]{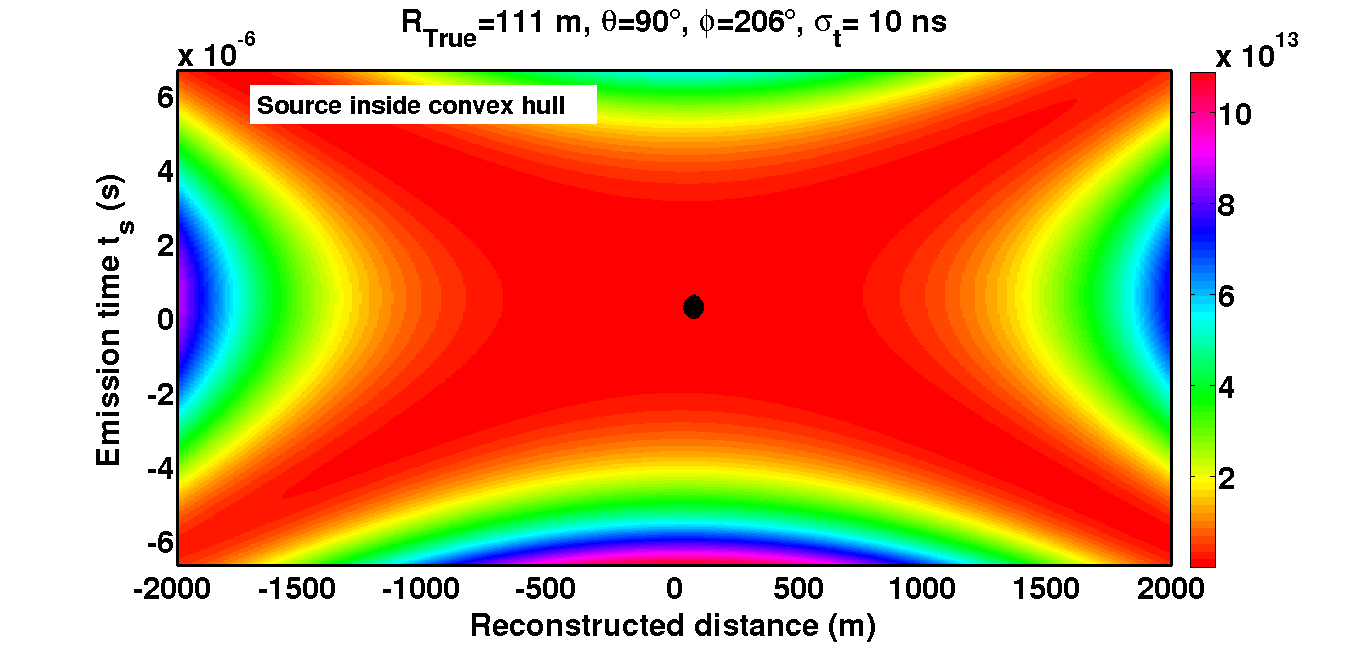}
\caption{Plots of the objective-function versus $R$ and versus the phase space
$(R,\, t)$, in the case of our testing array (2D), for a source on
the ground and located inside the convex surface of the antenna array.
This configuration leads to a single solution. In this case the problem
is well-posed. }
\label{obj_func_inside}
\end{figure}

\begin{figure}
\centering
\includegraphics[width=6cm,height=6cm]{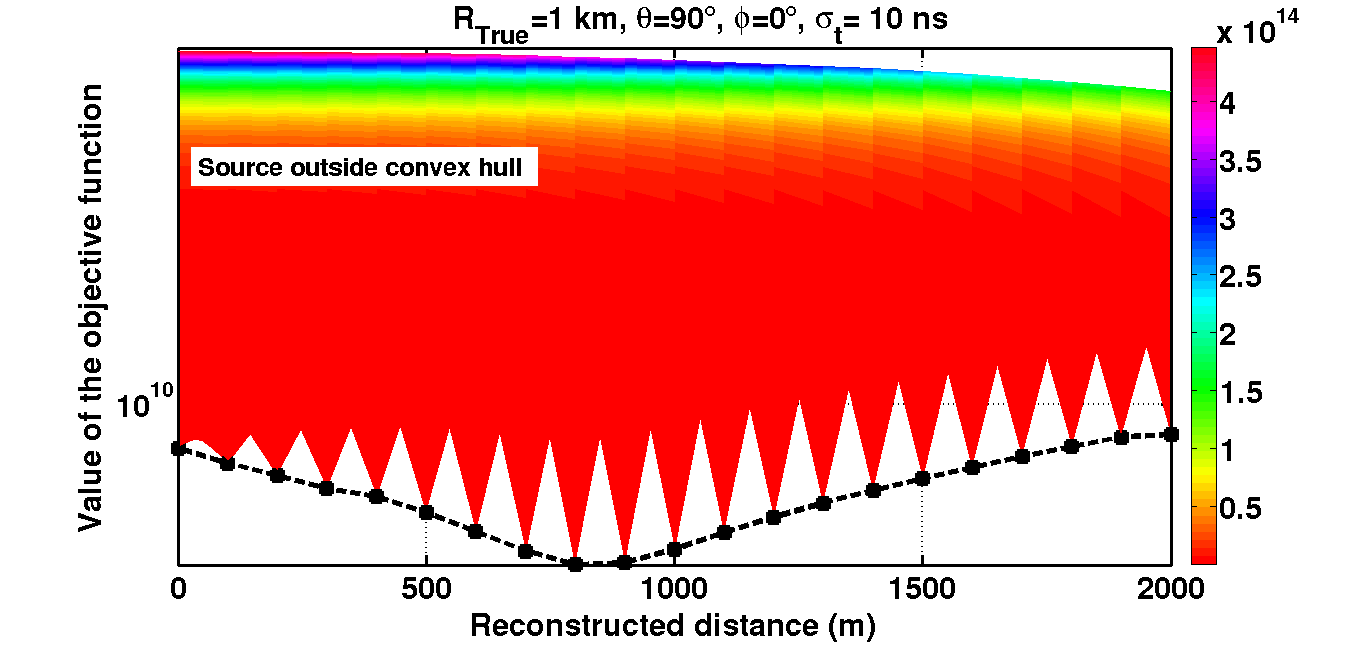}
\includegraphics[width=6cm,height=6cm]{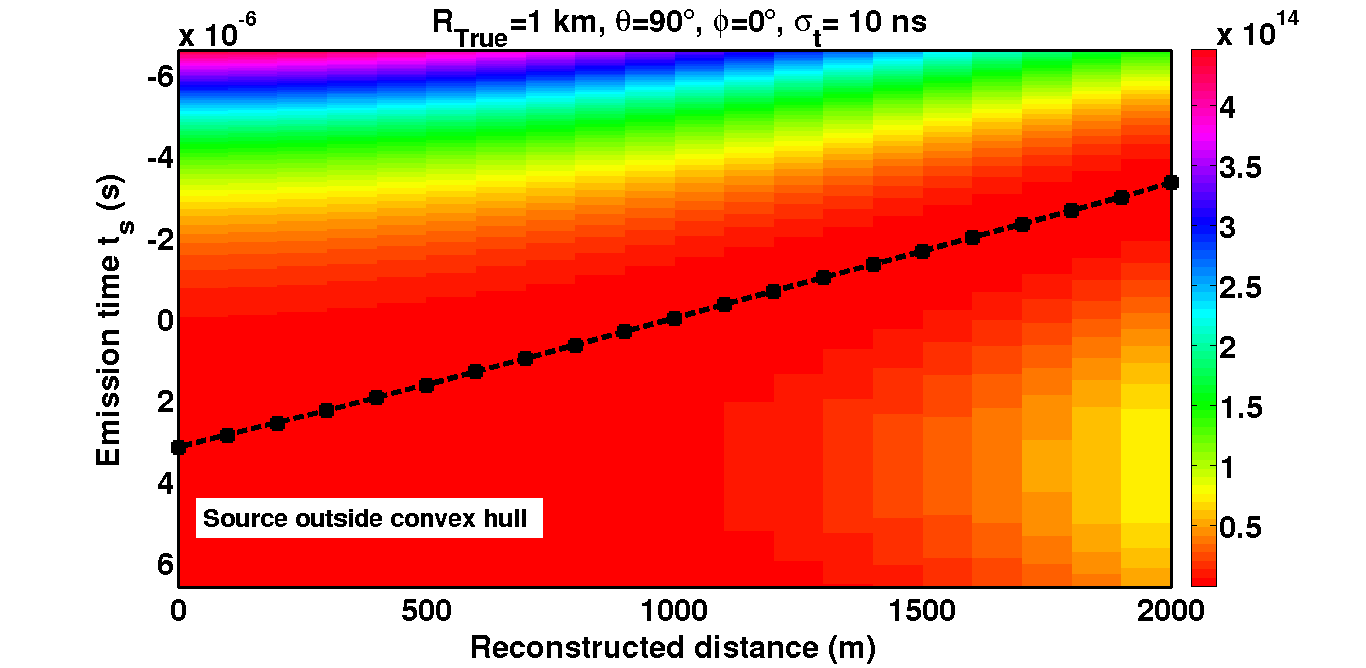}
\caption{Plots of the objective-function versus $R$ and versus the phase space
$(R,\, t)$, in the case of our testing array (2D) for a source outside
the convex hull. This configuration leads to multiple local minima.
All minima are located on the line joining the antenna barycenter
to the true source. In this case the problem is ill-posed. }
\label{obj_func_outside}
\end{figure}

The generalization of these results to real practical experience (with
a source located anywhere in the sky) was guided by our experimental
observations (performed through minimization algorithms) that provide
a first idea of what happens. For this, we chose to directly calculate
numerically the objective function for both general topologies: a
source inside the antenna array (ie. and at ground level) and an external
source to the antenna array (in the sky ). As can be deduced from
the results (see Figs. \ref{obj_func_inside} and \ref{obj_func_outside}),
for a surface antenna array, the convex hull is the surface defined
by the antennas illuminated. (An extrapolation of reasoning to a 3D
network (such as Ice Cube, ANTARES,...) should lead, this time, to
the convex volume of the setup).

Our results suggest the following interpretations:
\begin{itemize}
\item If the source is in the convex hull of the detector, the solution
is unique. In contrast, the location of the source outside the convex
hull of the detector, causes degeneration of solutions (multiple local
minimums) regarding to the constrained optimization problem. The source
position, outside or inside the array, affects the convergence of
reconstruction algorithms. 
\end{itemize}

\section{Proposed method of reconstruction }

Taking into account of the previous results, in order to avoid the
trap of the local minima with common algorithms, we chose to compute
directly the values of the objective function on a grid, using a subset
of phase space in the vicinity of the solution a priori, and assuming
that the minimum of the objective function correspond to the best
estimate of the position of the source of emission. The input parameters
are set from the planar fit, which provides both windows in $\theta$
and $\phi$. By cons, this method, being based on the search of the
minimum of the objective function using a grid calculation, the choice
of the metric becomes crucial. On the zenith and azimuth angles, the
metric can be adequately inferred from the value of the angular resolution
obtained by the current detectors, or $0.1^{\text{\textdegree}}$
for $\theta$ and $\phi$. Looking at the space metrics, the latter
can be inferred by considering the quantity $(c^{2}\sigma_{t}^{2})^{2}$,
ie. around one meter. The direction-priori is given by the planar
fit, while the quantity $r_{s}$ is left free in the range $0.1-20\: km$
(the upper bound being determined by the value of the curvature exploitable,
given the temporal resolutions currently available). 

\begin{figure}
\centering
\includegraphics[width=11cm,height=6cm]{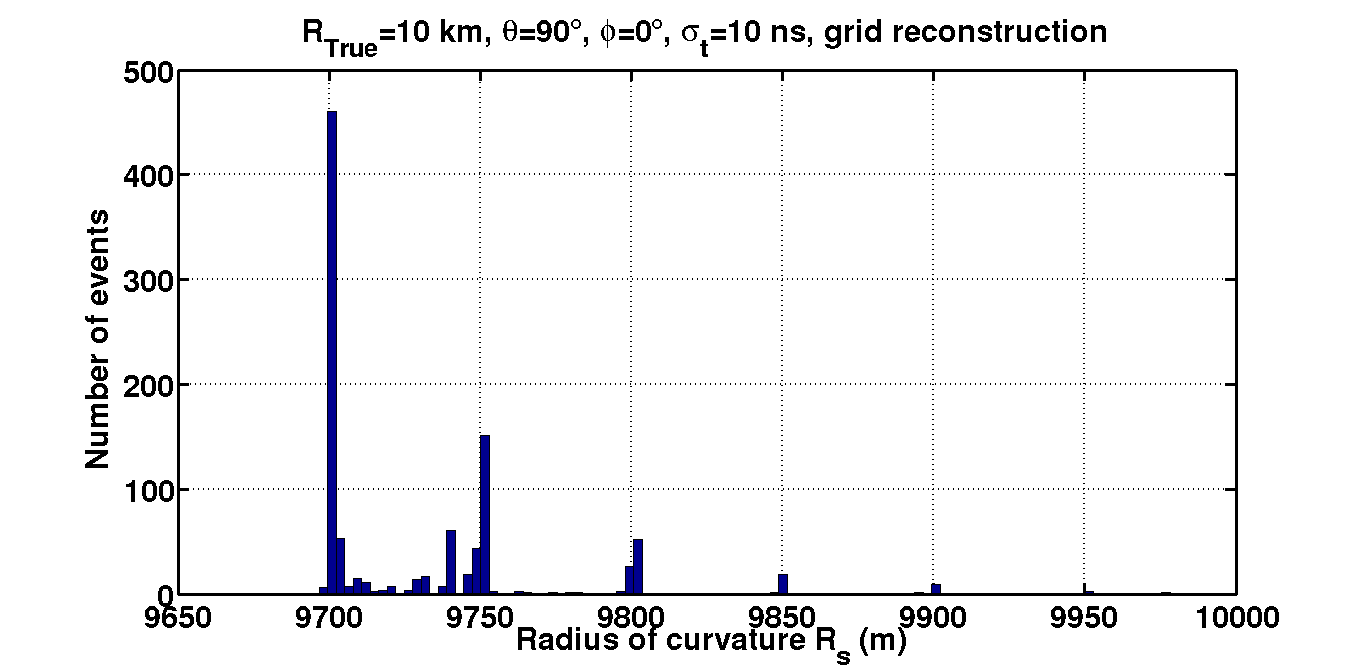}
\caption{Histogram of the reconstruction of a source located at 10 km from
the detector array, and using the grid method associated to the search
of the global minimum for each event. }
\label{Mindemin}
\end{figure}

A typical result obtained with our method is presented in Fig. \ref{Mindemin}
and a summary of the reconstructed parameters is given in table \ref{TabMindemin}
which have to be compared to those presented in the table \ref{ResulSimul}. 

\begin{table}
\centering %
\begin{tabular}{|l|l|l|l|l|}
\hline 
\multicolumn{5}{|c|}{Reconstruction results}\tabularnewline
\hline 
$\sigma_{t}(ns)$  & $R_{true}(m)$  & $R_{mean}(m)$  & $R_{mode}(m)$  & $\sigma^{R}(m)$ \tabularnewline
\hline 
$0$  & 1000  & 1000  & 1000  & 0 \tabularnewline
 & 3000  & 3000  & 3000  & 0 \tabularnewline
 & 10000  & 10000  & 10000  & 0 \tabularnewline
\hline 
3  & 1000  & 1010  & 998  & 58 \tabularnewline
 & 3000  & 2964  & 2700  & 214 \tabularnewline
 & 10000  & 9806  & 9700  & 161 \tabularnewline
\hline 
$10$  & 1000  & 987  & 902  & 150 \tabularnewline
 & 3000  & 2780  & 2700  & 149 \tabularnewline
 & 10000  & 9734  & 9700  & 50 \tabularnewline
\hline 
\end{tabular}\label{TabMindemin}\centering\caption{Summary of parameters reconstructed with our method for the same input
values as given in table 3. }
\end{table}

\section{Conclusion}

Experimental results indicated that the common methods of minimization
of spherical wavefronts could induce a mis-localisation of the emission
sources. In the current form of our objective function, a first elementary
mathematical study indicates that the source localization method may
lead to ill-posed problems, according to the actual source position.
To overcome this difficulty, we developed a simple method, based on
grid calculation of the objective function. This approach appears
to provide, at worst, an estimate as good as for the common algorithms
for locating the main point of the emission source, keeping in mind
that this method is not optimal in the sense of optimization theories.
However, further developments are without any doubt still necessary,
maybe based on advanced statistical theories, like for instance by
adding further information (as the signal amplitude or the functional
of the radio lateral distribution). This could be achieved by trying
a generalized objective function which includes these parameters.
In addition, the interactions with other disciplines which face this
problem could also provide tracks of work (especially regarding earth
sciences which focus on technics of petroleum prospecting).

\section{Appendix 1}

\subsection{Symbolic calculus}

Keeping the same notation as in table \ref{table_nomenclature}, the
objective function can be written: 
\[
f\left(X_{s}\right)=\frac{1}{2}\sum_{i=1}^{N}f_{i}^{2}\left(X_{s}\right)
\]
with $f_{i}\left(X_{s}\right)=\left(X_{s}-X_{i}\right)^{T}\cdot M\cdot\left(X_{s}-X_{i}\right)=\left\Vert \overrightarrow{r_{s}}-\overrightarrow{r_{i}}\right\Vert ^{2}-\left(t_{s}^{*}-t_{i}^{*}\right)^{2}$. 

The formula $\nabla f\left(X_{s}\right)=\sum f_{i}\left(X_{s}\right)\cdot\nabla f_{i}\left(X_{s}\right)$
is derived from the formula of a product derivation. Using the bi-linearity
of the inner product, we show that $\nabla f_{i}\left(X_{s}\right)=2M\cdot\left(X_{s}-X_{i}\right)$.
By injecting this formula into the formula of $\nabla f$ , we obtain
the following formula: 
\begin{align*}
\nabla f\left(X_{s}\right)= & \sum f_{i}\left(X_{s}\right)\cdot\nabla f_{i}\left(X_{s}\right)\\
= & \sum f_{i}\left(X_{s}\right)\cdot2M\cdot\left(X_{s}-X_{i}\right)
\end{align*}
It then leads to the following form: 
\[
\frac{1}{2}\nabla f\left(X_{s}\right)=\left(\sum f_{i}\left(X_{s}\right)\right)M\cdot X_{s}-M\cdot\left(\sum f_{i}\left(X_{s}\right)X_{i}\right)
\]
 With the same method, the second derivative matrix (Hessian matrix)
is given by the following formula :
\[
\nabla^{2}f\left(X_{s}\right)=\sum\nabla f_{i}\left(X_{s}\right)\cdot\nabla f_{i}\left(X_{s}\right)^{T}+\sum f_{i}\left(X_{s}\right)\cdot\nabla^{2}f_{i}\left(X_{s}\right)
\]

By injecting in the previous formula the following formula of the
second derivatives $\nabla^{2}f_{i}\left(X_{s}\right)=2M$ and by
using the relation $\left(AB\right)^{T}=B^{T}A^{T}$, we get the following
formula: 
\begin{align*}
\nabla^{2}f\left(X_{s}\right)= & \sum\nabla f_{i}\left(X_{s}\right)\cdot\nabla f_{i}\left(X_{s}\right)^{T}+\sum f_{i}\left(X_{s}\right)\cdot\nabla^{2}f_{i}\left(X_{s}\right)\\
= & \sum2M\cdot\left(X_{s}-X_{i}\right)\cdot\left(2M\cdot\left(X_{s}-X_{i}\right)\right)^{T}+\sum f_{i}\left(X_{s}\right)\cdot2M\\
= & 4M\cdot\left[\sum\left(X_{s}X_{s}^{T}-X_{s}X_{i}^{T}-X_{i}X_{s}^{T}+X_{i}X_{i}^{T}\right)\right]\cdot M+2\left(\sum f_{i}\left(X\right)\right)\cdot M\\
= & 4M\cdot\left[NX_{s}X_{s}^{T}+\sum X_{i}X_{i}^{T}-X_{s}\left(\sum X_{i}\right)^{T}-\left(\sum X_{i}\right)X_{s}^{T}\right]\cdot M+2\left(\sum f_{i}\left(X_{s}\right)\right)\cdot M
\end{align*}
Both relationships correspond to the end-calculus forms given in the
$3^{\textrm{rd}}$ section. These forms are easy to handle for symbolic
calculus but not convenient for explicit calculation used for studying
the convexity.

\subsection{Taylor expansion and explicit calculus}

An explicit form for the objective function first and second differential
can be obtained using a Taylor expansion. Indeed, the function$f$
is an element of $C^{\infty}\left(\mathbb{R}^{4},\mathbb{R}\right)$
\footnote{The function is also an element of the algebra $\mathbb{R}\left[X_{1},\ldots,X_{4}\right]$%
} and is therefore differentiable in the sense of \emph{Fréchet}.

Let $X_{s}=\left(\overrightarrow{r_{s}},t_{s}^{*}\right)^{T}$ be
a fixed vector of $\mathbb{R}^{4}$ and $\overrightarrow{\varepsilon}=\left(\overrightarrow{h},t^{*}\right)^{T}$
another vector of $\mathbb{R}^{4}$. In order to simplify the calculus,
we use the following notations :

$K_{i}=\left\Vert \overrightarrow{r_{s}}-\overrightarrow{r_{i}}\right\Vert _{2}^{2}-\left(t_{s}^{*}-t_{i}^{*}\right)^{2}$
a constant term when setting the vector $X_{s}$;

$L_{i}\left(\overrightarrow{\varepsilon}\right)=\left\langle \overrightarrow{r_{s}}-\overrightarrow{r_{i}}\mid\overrightarrow{h}\right\rangle -\left(t_{s}^{*}-t_{i}^{*}\right)\cdot t^{*}$
the linear form;

and $Q\left(\overrightarrow{h},t^{*}\right)=\left\Vert \overrightarrow{h}\right\Vert _{2}^{2}-t^{*2}$
the quadratic form.

The Taylor expansion leads to: 
\begin{align*}
f\left(X_{s}+\overrightarrow{\varepsilon}\right)= & \frac{1}{2}\sum_{i}\left(\left\Vert \overrightarrow{r_{s}}+\overrightarrow{h}-\overrightarrow{r_{i}}\right\Vert _{2}^{2}-\left(t_{0}^{*}+t^{*}-t_{i}^{*}\right)^{2}\right)^{2}\\
= & \frac{1}{2}\sum_{i}\left(\left\langle \overrightarrow{r_{s}}+\overrightarrow{h}-\overrightarrow{r_{i}}\mid\overrightarrow{r_{s}}+\overrightarrow{h}-\overrightarrow{r_{i}}\right\rangle -\left(t_{s}^{*}+t^{*}-t_{i}^{*}\right)^{2}\right)^{2}\\
= & \frac{1}{2}\sum_{i}\left(\left\Vert \overrightarrow{r_{s}}-\overrightarrow{r_{i}}\right\Vert _{2}^{2}+\left\Vert \overrightarrow{h}\right\Vert _{2}^{2}+2\left\langle \overrightarrow{r_{s}}-\overrightarrow{r_{i}}\mid\overrightarrow{h}\right\rangle -\left(t_{s}^{*}-t_{i}^{*}\right)^{2}-t^{*2}-2t^{*}\left(t_{s}^{*}-t_{i}^{*}\right)\right)^{2}
\end{align*}

Using the multinomial expansion, the function $f$ can then be approximated
by the second-order Taylor expansion following: 
\[
f\left(\overrightarrow{r_{s}}+\overrightarrow{h},t_{s}^{*}+t^{*}\right)\approx\frac{1}{2}\sum_{i}K_{i}^{2}+2\sum_{i}K_{i}\cdot L_{i}\left(\overrightarrow{h},t^{*}\right)+2\sum_{i}L_{i}^{2}\left(\overrightarrow{h},t^{*}\right)+\left(\sum_{i}K_{i}\right)\cdot Q\left(\overrightarrow{h},t^{*}\right)
\]

We identify from this formula:

the constant term $\frac{1}{2}\underset{i}{\sum}K_{i}^{2}$;

the linear term which is $\nabla f\left(X_{s}\right)^{T}\cdot\overrightarrow{\varepsilon}=\left[2\cdot\underset{i}{\sum}K_{i}\left(\begin{array}{c}
\overrightarrow{r_{s}}-\overrightarrow{r_{i}}\\
t_{i}^{*}-t_{s}^{*}
\end{array}\right)\right]^{T}\cdot\overrightarrow{\varepsilon}$ (the $f$ first differential in $\left(\overrightarrow{r_{s}},t_{s}^{*}\right)$
);

and the quadratic form at the point $X_{s}$: {\tiny 
\[
\frac{1}{2}Q\left(X_{s},X_{i}\right)=\left[\begin{array}{cccc}
\sum_{i}K_{i}+2\sum_{i}\left(x_{s}-x_{i}\right)^{2} & 2\sum_{i}\left(x_{s}-x_{i}\right)\left(y_{s}-y_{i}\right) & 2\sum_{i}\left(x_{s}-x_{i}\right)\left(z_{s}-z_{i}\right) & 2\sum_{i}\left(x_{s}-x_{i}\right)\left(t_{i}^{*}-t_{s}^{*}\right)\\
* & \sum_{i}K_{i}+2\sum_{i}\left(y_{s}-y_{i}\right)^{2} & 2\sum_{i}\left(y_{s}-y_{i}\right)\left(z_{s}-z_{i}\right) & 2\sum_{i}\left(y_{s}-y_{i}\right)\left(t_{i}^{*}-t_{s}^{*}\right)\\
* & * & \sum_{i}K_{i}+2\sum_{i}\left(z_{s}-z_{i}\right)^{2} & 2\sum_{i}\left(z_{s}-z_{i}\right)\left(t_{i}^{*}-t_{s}^{*}\right)\\
* & * & * & -\sum_{i}K_{i}+2\sum_{i}\left(t_{i}^{*}-t_{s}^{*}\right)^{2}
\end{array}\right]
\]
}{\tiny \par}

which is the $f$ Hessian matrix in $\left(\overrightarrow{r_{s}},t_{s}^{*}\right)$,
or the second differential of $f$ also denoted $\nabla^{2}f\left(X_{s},X_{i}\right)$.
The use of $*$ indicates that the coefficients above and below the
diagonal are equal (\emph{Schwarz} Lemma). The quadratic form represented
by this matrix gives us the local second-order properties for the
function $f$. To show that a critical point is a local minimum, it
will suffice to verify that the Hessian matrix is definite positive
in the vicinity of this point.

\subsection{Convexity property}

The convex analysis occupies a capital place in the problems of minimization.
Indeed, an important theorem yet intuitive stated that if a convex
function has a local minimum, it is automatically global. We will
shows that the function $f$ is not convex in $\mathbb{R}^{4}$, i.e.
that the Hessian matrix in non-positive define. 

Let $\nabla^{2}f\left(X\right)$ the Hessian matrix, and let's suppose
$d$ a vector, since the function $f$ is twice differentiable, using
the Sylvester's criterion \cite{key-13} to characterize the convexity
of $f$ , we can write the following equivalence: 
\[
f\;\textrm{is convex}\Leftrightarrow\textrm{ Hessian is positive semi-definite}\Leftrightarrow\textrm{ All Hessian principal minors are just nonnegative}
\]
\[
f\;\textrm{is convex}\Leftrightarrow\forall d,\:\forall X,\; d^{T}\cdot\nabla^{2}f\left(X\right)\cdot d\geqslant0
\]
So if we can find an element $X$ and $d$ such as $d^{T}\cdot\nabla^{2}f\left(X\right)\cdot d<0$,
$f$ will be non-positive definite. For this, it is sufficient to
find a single negative principal minor to demonstrate the Hessian
matrix is non-positive definite. The objective function f will present
then several local minimums and will be thus locally non-convex.

So let $Q$ the explicit expression of the Hessian and let us choose
$d^{T}\:=\:(0\:0\:0\:1)$ then: 
\[
d^{T}\cdot\nabla^{2}f(X)\cdot d=(0\:0\:0\:1)\cdot Q(X_{s},X_{i})\cdot\left(\begin{array}{c}
0\\
0\\
0\\
1
\end{array}\right)
\]
\[
=-\sum_{i}K_{i}+2\sum_{i}\left(t_{i}^{*}-t_{s}^{*}\right)^{2}
\]
which is represent the principal minor of order $4$ of the Hessian. 

For a family of fixed positions antennas and for a signal source with
coordinates $X_{s}$ such as $y_{s}=z_{s}=t_{s}^{*}=0$, the negativity
condition of the principal minor of order 4 can then written: 
\[
\sum_{i}\left(x_{s}-x_{i}\right)^{2}>\sum_{i}\left(-y_{i}^{2}-z_{i}^{2}+3t_{i}^{*2}\right)
\]
Now the left term tends to infinity when the source tends to infinity%
\footnote{We say that the function is \emph{coercive}%
}. It is written in terms of limits, 
\[
\lim_{\left|x_{s}\right|\rightarrow+\infty}\sum_{i}\left(x_{s}-x_{i}\right)^{2}=+\infty\Leftrightarrow\forall A>0,\:\exists\eta>0\diagup\left|x_{s}\right|>\eta\Rightarrow\sum_{i}\left(x_{s}-x_{i}\right)^{2}>A
\]
Taking a value $\underset{i}{\sum}\left(-y_{i}^{2}-z_{i}^{2}+3t_{i}^{2*}\right)$
of the constant $A$, it exist a real $\eta$ and therefore a $x_{s}$
such that $\underset{i}{\sum}\left(x_{s}-x_{i}\right)^{2}>\underset{i}{\sum}\left(-y_{i}^{2}-z_{i}^{2}+3t_{i}^{*2}\right)$.
We deduce that the function is not convex in the vicinity of this
point. It suffices to take $d^{T}=(0\:0\:0\:1)$ and $x_{s}=\eta+1$.

\section{Appendix 2}

\subsection{Degeneration line for a linear antenna array}

According to experimental data analysis and to our simulations (see
Fig. \ref{codanoise} and \ref{obj_func_outside}), the results of
the common minimization algorithms appear to fall on a half-line in
the phase space $(x,y,z)$ which we shall call the degeneration line,
which is linked to the existence of local minima. We present the mathematical
development in the case of a linear array using an analysis-synthesis
method. Then we try to generalize results to the higher dimension
cases. 

Let suppose $X_{s}=(x_{s},t_{s}^{*})$ a critical point of $f$, ie.
$\nabla f(X_{s})=0$, for a linear array, the minimization problem
with constraints can written: 

\[
\left\{ \begin{aligned}arg~min~f(x_{s},t_{s}^{*})=\frac{1}{2}\sum_{i=1}^{N}((x_{s}-x_{i})^{2}-(t_{s}^{*}-t_{i}^{*})^{2})^{2}~~1\leqslant i\leqslant N\\
Propagation~constraint:~|x_{s}-x_{i}|=|t_{s}^{*}-t_{i}^{*}|\\
Causality~constraint:~t_{s}^{*}<min_{i}(t_{i}^{*})
\end{aligned}
\right.
\]

and Let suppose $L=\left(\begin{array}{c}
L\\
L
\end{array}\right)$ so that $X_{s}-L$ is also a a solution of the minimization problem,
ie. $\nabla f(X_{s}-L)=0$)

The Jacobian of $f$ is written as: 
\[
\vec{\nabla}f\left(x_{s},t^{*}\right)=2\left(\begin{array}{c}
\underset{i}{\sum}\left(x_{s}-x_{i}\right)\left(\left(x_{s}-x_{i}\right)^{2}-\left(t_{s}^{*}-t_{i}^{*}\right)^{2}\right)\\
\underset{i}{\sum}\left(t_{i}^{*}-t_{s}^{*}\right)\left(\left(x_{s}-x_{i}\right)^{2}-\left(t_{s}^{*}-t_{i}^{*}\right)^{2}\right)
\end{array}\right)
\]
If $X_{s}$ being a critical point, this leads to two equations: 
\[
\begin{cases}
\sum_{i}\left(x_{s}-x_{i}\right)\left(\left(x_{s}-x_{i}\right)^{2}-\left(t_{s}^{*}-t_{i}^{*}\right)^{2}\right)=0\quad\left(1\right)\\
\sum_{i}\left(t_{i}^{*}-t_{s}^{*}\right)\left(\left(x_{s}-x_{i}\right)^{2}-\left(t_{s}^{*}-t_{i}^{*}\right)^{2}\right)=0\quad\left(2\right)
\end{cases}
\]
Assuming that $X_{s}-L$ being also a critical point, this leads to
two equations: 
\[
\begin{cases}
\sum_{i}\left(x_{s}-x_{i}-L\right)\left(\left(x_{s}-x_{i}-L\right)^{2}-\left(t_{s}^{*}-t_{i}^{*}-L\right)^{2}\right)=0\quad\left(3\right)\\
\sum_{i}\left(t_{i}^{*}-t_{s}^{*}+L\right)\left(\left(x_{s}-x_{i}-L\right)^{2}-\left(t_{s}^{*}-t_{i}^{*}-L\right)^{2}\right)=0\quad\left(4\right)
\end{cases}
\]
By developing the equation $\left(3\right)$ and by using the equation
$\left(1\right)$, then: 
\[
\left(3\right)\Rightarrow~~\sum_{i}\left(x_{s}-x_{i}\right)\left[\left(x_{s}-x_{i}\right)^{2}-\left(t_{s}^{*}-t_{i}^{*}\right)^{2}-2L\left[\left(x_{s}-x_{i}\right)-\left(t_{s}^{*}-t_{i}^{*}\right)\right]\right]-L\sum_{i}\left(x_{s}-x_{i}\right)^{2}-\left(t_{s}^{*}-t_{i}^{*}\right)^{2}
\]
\[
\ldots~~+2L^{2}\sum_{i}\left[\left(x_{s}-x_{i}\right)-\left(t_{s}^{*}-t_{i}^{*}\right)\right]=0
\]

\[
\Rightarrow~~-L\sum_{i}\left(x_{s}-x_{i}\right)^{2}+L^{2}\sum_{i}\left[\left(x_{s}-x_{i}\right)-\left(t_{s}^{*}-t_{i}^{*}\right)\right]-L\sum_{i}\left(x_{s}-x_{i}\right)^{2}-\left(t_{s}^{*}-t_{i}^{*}\right)^{2}
\]
\[
\ldots~~+L\sum_{i}\left(x_{s}-x_{i}\right)\left(t_{s}^{*}-t_{i}^{*}\right)=0
\]
The set of constraints requires that the term $\underset{i}{\sum}\left(x_{s}-x_{i}\right)^{2}-\left(t_{s}^{*}-t_{i}^{*}\right)^{2}$
is null. We get the simplified equation: 
\[
L\sum_{i}\left(x_{s}-x_{i}\right)-\left(t_{s}^{*}-t_{i}^{*}\right)=\sum_{i}\left(x_{s}-x_{i}\right)\left(\left(x_{s}-x_{i}\right)-\left(t_{s}^{*}-t_{i}^{*}\right)\right)
\]
In the cases where $x_{s}-x_{i}<0$ for all $i$, the set of constraints
is equivalent to $\left(x_{s}-x_{i}\right)-\left(t_{s}^{*}-t_{i}^{*}\right)=0$.
Thus, if one assumes that $x_{s}-x_{i}<0$ for all $i$, i.e that
the source is outside the array convex hull (a segment), we find that
previous implications are equivalences and thus that equation $\left(3\right)$
is verified. Operating in the same manner for the equation $\left(4\right)$,
we obtain the following equations: 
\[
\left(4\right)\Rightarrow\sum_{i}\left(t_{i}^{*}-t_{s}^{*}\right)\left[\left(x_{s}-x_{i}\right)^{2}-\left(t_{s}^{*}-t_{i}^{*}\right)^{2}-2L\left[\left(x_{s}-x_{i}\right)-\left(t_{s}^{*}-t_{i}^{*}\right)\right]\right]+L\sum_{i}\left(x_{s}-x_{i}\right)^{2}-\left(t_{s}^{*}-t_{i}^{*}\right)^{2}
\]

\[
\ldots-2L^{2}\sum_{i}\left[\left(x_{s}-x_{i}\right)-\left(t_{s}^{*}-t_{i}^{*}\right)\right]=0
\]

\[
\Rightarrow-2L\sum_{i}\left(t_{i}^{*}-t_{s}^{*}\right)\left(x_{s}-x_{i}\right)-2L\sum_{i}\left(t_{i}^{*}-t_{s}^{*}\right)^{2}+L\sum_{i}\left(x_{s}-x_{i}\right)^{2}-\left(t_{s}^{*}-t_{i}^{*}\right)^{2}
\]
\[
\ldots-2L^{2}\sum_{i}\left(x_{s}-x_{i}\right)-\left(t_{s}^{*}-t_{i}^{*}\right)=0
\]
Using the set of constraints as above, we obtain the following equation:
\[
L\sum_{i}\left(x_{s}-x_{i}\right)-\left(t_{s}^{*}-t_{i}^{*}\right)=\sum_{i}\left(t_{i}^{*}-t_{s}^{*}\right)\left(\left(x_{s}-x_{i}\right)-\left(t_{s}^{*}-t_{i}^{*}\right)\right)
\]
The same analysis as above gives us the condition that the source
is out of the antennas convex hull. This degeneration is an important
point because it determines the convergence of minimization algorithms.
In this case the problem of the reconstruction is ill-posed. \bigskip{}

The generalization of the previous calculation to higher dimensions
is more delicate, insofar as there are infinitely many directions
in which the source can move. The idea now is to translate the source,
from its position $\overrightarrow{r_{s}}$, simultaneously in all
directions $\overrightarrow{r_{s}}-\overrightarrow{r_{i}}$ and with
the same distances. We define the unit vector on the direction source-antenna.
It will be noted: $\overrightarrow{e_{i}}=\frac{\overrightarrow{r_{s}}-\overrightarrow{r_{i}}}{\left\Vert \overrightarrow{r_{s}}-\overrightarrow{r_{i}}\right\Vert _{2}}$.
The translation spatial direction thus defined, is given by the vector
$\overrightarrow{L}=\underset{i}{\sum}\overrightarrow{e_{i}}=\underset{i}{\sum}\frac{\overrightarrow{r_{s}}-\overrightarrow{r_{i}}}{\left\Vert \overrightarrow{r_{i}}-\overrightarrow{r_{s}}\right\Vert _{2}}=-\underset{i}{\sum}\frac{\overrightarrow{r_{i}}}{\left\Vert \overrightarrow{r_{s}}-\overrightarrow{r_{i}}\right\Vert _{2}}+\left(\underset{i}{\sum}\frac{1}{\left\Vert \overrightarrow{r_{s}}-\overrightarrow{r_{i}}\right\Vert _{2}}\right)\overrightarrow{r_{s}}$.
Considering the reduced temporal variables, the wave required delay
to traverse the distance induced by the translation $\left\Vert \overrightarrow{L}\right\Vert $.
Let $V$ the vector of coordinates $V=\left(\overrightarrow{L},\left\Vert \overrightarrow{L}\right\Vert \right)^{T}$.
We write the first order optimality condition for the vector of $\mathbb{\mathbb{R}}{}^{4}$:
$X_{s}-V$:
\begin{eqnarray*}
\nabla f\left(X_{s}-V\right)= & \left(\sum f_{i}\left(X_{s}-V\right)\right)\cdot M\cdot\left(X_{s}-V\right)-M\cdot\left(\sum f_{i}\left(X_{s}-V\right)\cdot X_{i}\right)
\end{eqnarray*}
By introducing the condition $\nabla f\left(X_{s}\right)=0$ which
implies that: 
\[
\left(\sum f_{i}\left(X_{s}\right)\right)\cdot M\cdot X_{s}-M\cdot\left(\sum f_{i}\left(X_{s}\right)X_{i}\right)=0
\]
we obtain: 
\begin{align*}
\nabla f\left(X_{s}-V\right)= & N\left(V^{T}\cdot M\cdot V\right)\cdot M\cdot\left[X_{s}-V-\frac{1}{N}\sum_{i}X_{i}\right]\\
\, & 2M\left(\sum\left(X_{s}-X_{i}\right)^{T}M\cdot V\cdot\overrightarrow{X_{i}}\right)-2\left(\sum\left(X_{s}-X_{i}\right)^{T}M\cdot V\right)\cdot M\cdot\left(X_{s}-V\right)\\
\, & -\left(\sum f_{i}\left(X_{s}\right)\right)\cdot M\cdot V
\end{align*}
According to the imposed form of the vector $\overrightarrow{V}$,
then: 
\[
V^{T}\cdot M\cdot V=\left(\overrightarrow{L}\:\left\Vert \overrightarrow{L}\right\Vert _{2}\right)M\left(\overrightarrow{L}\:\left\Vert \overrightarrow{L}\right\Vert _{2}\right)^{T}=0
\]
It remains then the following expression: 
\begin{align*}
\nabla f\left(X_{s}-V\right)= & 2M\left(\sum\left(X_{s}-X_{i}\right)^{T}M\cdot V\cdot X_{i}\right)-2\left(\sum\left(X_{s}-X_{i}\right)^{T}M\cdot V\right)\cdot M\cdot\left(X_{s}-V\right)\\
\, & -\left(\sum f_{i}\left(X_{s}\right)\right)\cdot M\cdot V
\end{align*}
The resolution of this equation should lead to an analytical expression
for the topology of critical points. We failed to develop it, but
we can already see that the explicit development leads to cross terms
that will make simplifications difficult. Therefore, we have tried
again an intuitive approach based on the numerical simulations presented
section 3.3.

\section{Appendix 3}

\subsection{Convex hull for a linear antenna array}

\begin{figure}[!h]
\centering
\includegraphics[width=11cm,height=6cm]{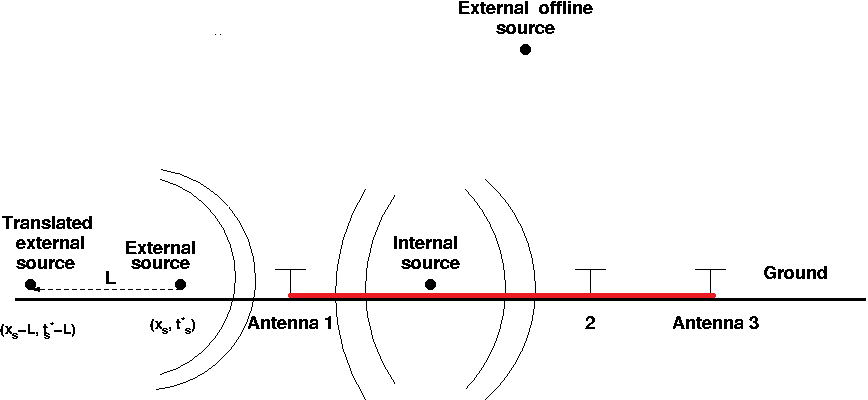}
\caption{Scheme of the reconstruction problem of spherical waves for a 1D array
of antennas. For this configuration, the convex hull is the segment
shown in red. }
\label{Pb_1D-1} 
\end{figure}

Let us consider the sub-array of the 3 aligned upper antennas presented
in Fig. \ref{how_test_estim}). The figure \ref{Pb_1D-1} shows the
physical principle of the reconstruction of the source.

\begin{figure}[!h]
\centering
\includegraphics[width=11cm,height=6cm]{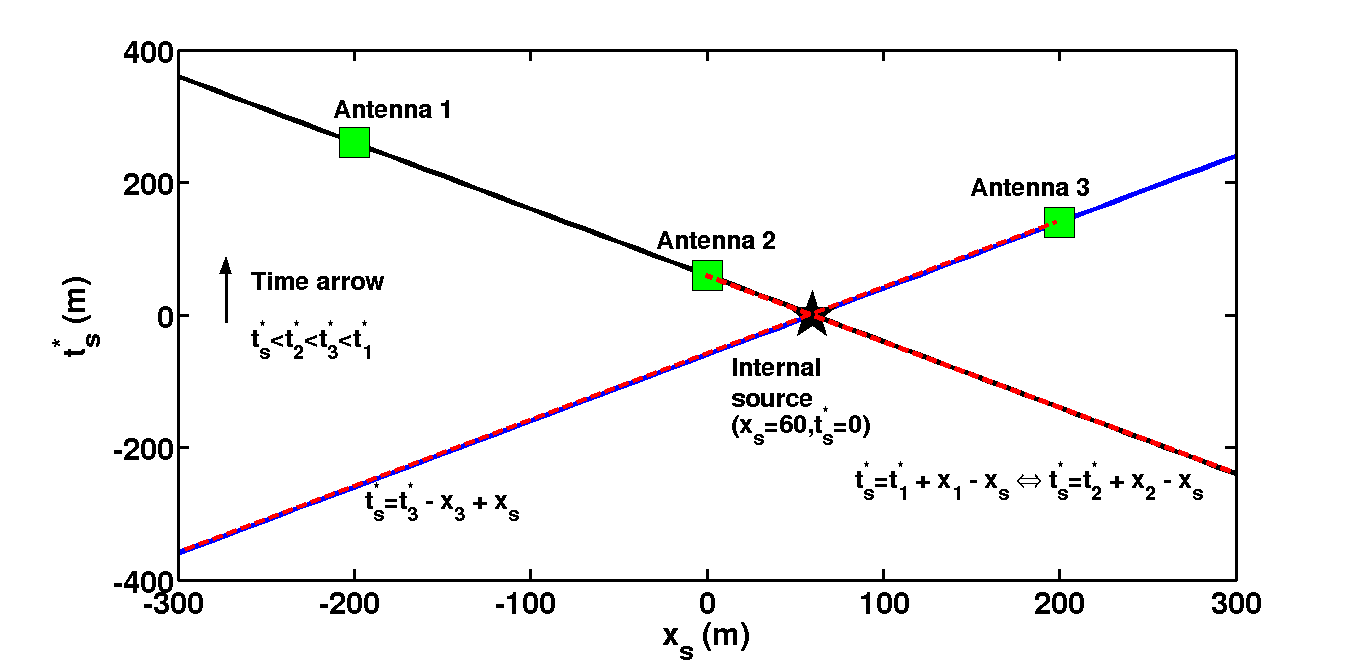}
\caption{Phase space representation in the case of a linear array of three
antennas (shown as green squares located at $x_{1}=-200\: m$, $x_{2}=0\: m$,
$x_{3}=200\: m$). The source is located at $x_{s}=60\, m$ when the
instant of the emission is taken as the time origin ($t_{s}=0\: s$).
Because the source is outside this sub-array, the constraints on the
positions of antenna 1 and 2 lead to the same equation $t_{s}=60-x_{s}$
(black line). Equation for the antenna 3 (blue line) leads to $t_{s}=-60+x_{s}$.
The causality conditions restrict the initial lines to two half-lines
(red lines). The source location (black star) is at the intercept
of the both half-lines.}
\label{Cstr_1D_inside} 
\end{figure}

Three situations must be considered: 
\begin{itemize}
\item the source located inside the array; 
\item the source located outside the array but on the detector axis; 
\item the source located outside this main axis. The latter corresponds
to the typical problems encountered with of the man-made emitters
located on the ground. 
\end{itemize}
The first situation leads to 2 half-lines cutting each other in a
single point: the solution is unique (Figure \ref{Cstr_1D_inside})
and the localization problem is well-posed. The source is unique and
inside the line segment linking the nearest antennas to the source.
This segment correspond to the convex hull within this geometry. We
can also note that only the two antennas flanking the source then
play a role in its localization. The problem writes:

\[
\left\{ \begin{aligned}arg~min~f(x_{s},t_{s}^{*})=\frac{1}{2}\sum_{i=1}^{N}((x_{s}-x_{i})^{2}-(t_{s}^{*}-t_{i}^{*})^{2})^{2}\\
Propagation~constraint:~|x_{s}-x_{i}|=|t_{s}^{*}-t_{i}^{*}|\\
Causality~constraint:~t_{s}^{*}<min_{i}(t_{i}^{*})
\end{aligned}
\right.
\]

\begin{figure}[!h]
\centering 
\includegraphics[width=11cm,height=6cm]{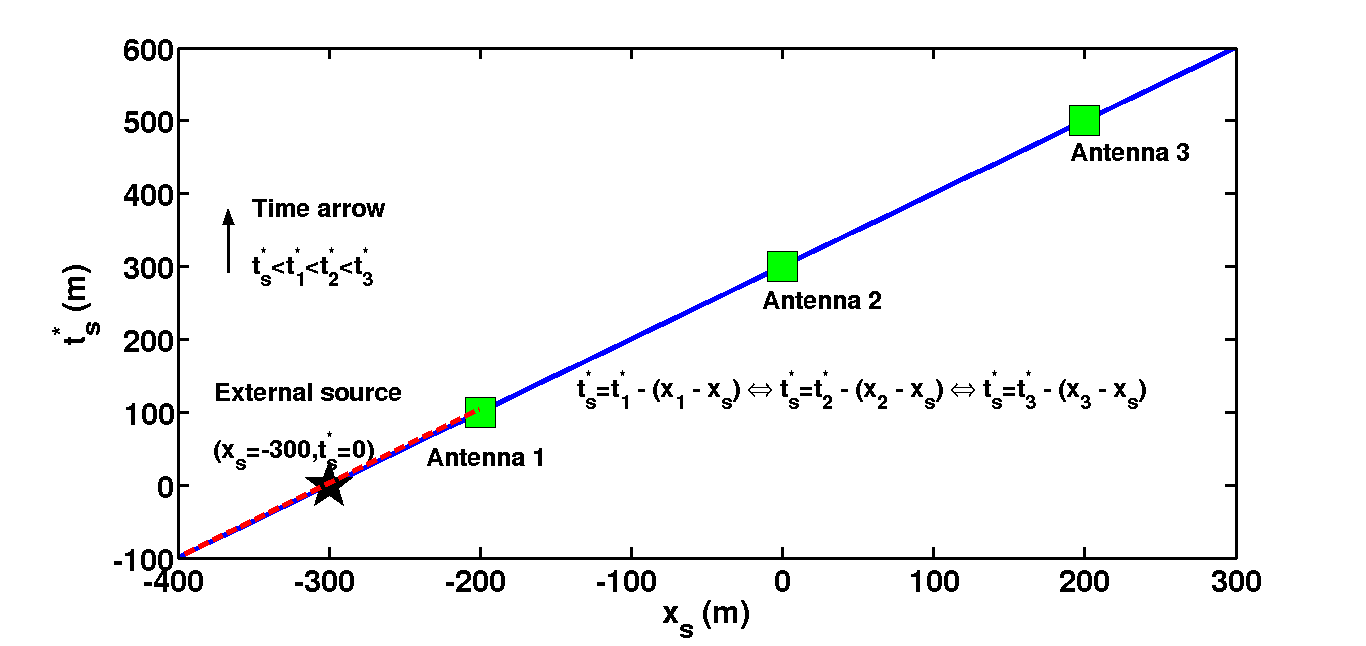}
\caption{Same as figure \ref{Cstr_1D_inside} but for an on-axis source outside
the linear array of three antennas. The whole constraints lead the
same equation $t_{s}^{*}=60-x_{s}$. All points belonging to the lower
half-line are solutions of the source localization problem (red dashed
line), which becomes, in this case, ill-posed, and creates the degenerations. }
\label{Cstr_1D_outside} 
\end{figure}

About the source on-axis, but outside the convex hull, the arrival
times between the antennas, are no longer related to the source position,
but to their locations. Whatever their positions, the time differences
remain constant (for equally spaced antennas). It becomes impossible
to distinguish between two different shifted sources by any length.
The only relevant information lies in the direction of propagation
of the wave (see figure \ref{Cstr_1D_outside}). This result appears
by a degeneration of solutions because all points located on the half-line
starting from the first tagged antenna are solutions of the problem
which is ill-posed. The source is outside the convex hull of the antenna
array. 

On the configuration where source located outside this antenna axis
(problem in two dimensions), the solving starts with:

\[
\left\{ \begin{aligned}arg~min~f(x_{s},y_{s},t_{s}^{*})=\frac{1}{2}\sum_{i=1}^{N}((x_{s}-x_{i})^{2}+y_{s}^{2}-(t_{s}^{*}-t_{i}^{*})^{2})^{2}\\
Propagation~constraint:~(x_{s}-x_{i})^{2}+y_{s}^{2}=(t_{s}^{*}-t_{i}^{*})^{2}\\
Causality~constraint:~t_{s}^{*}<min_{i}(t_{i}^{*})
\end{aligned}
\right.
\]

\begin{figure}[!h]
\centering 
\includegraphics[width=10cm,height=7cm]{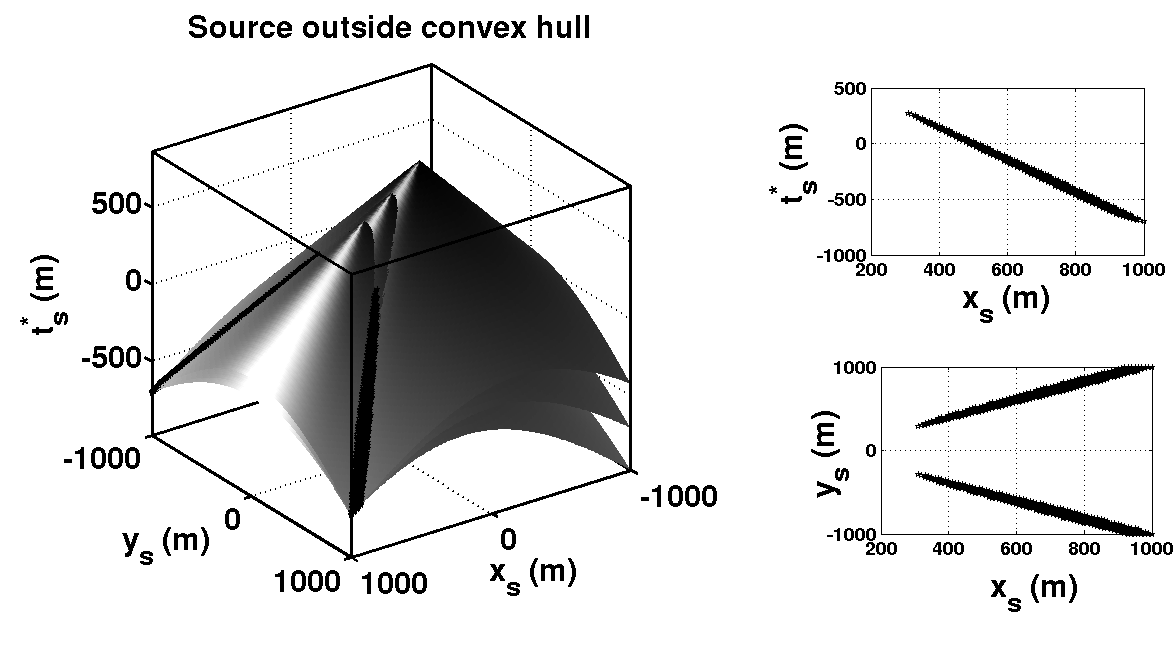}
\caption{Reconstruction problem of spherical waves for a 1D array of antennas
with the source outside the convex hull. The local minima are located
at the intersection of the cones. }
\label{Pb_2D} 
\end{figure}

The constraint set reduces the problem of characterization of critical
points to the search of the half-cones intersections induced by each
antenna, in the 3 dimensional phase space $(x,\, y\,,\, t)$ and which
presents a great similarity of constraints with the light cone used
in special relativity (Fig. \ref{Pb_2D}). Intersection of the half-cones,
two to two, induces multiple critical points which are local minima.

\end{document}